\documentclass[letterpaper,twocolumn,10pt]{article}
\usepackage{usenix2019_v3}

\usepackage{tikz}
\usepackage{amsmath}

\usepackage{filecontents}

\usepackage{authblk}

\usepackage{enumitem}
\usepackage{multirow}
\usepackage{tabularx}
\usepackage[group-separator={,}]{siunitx}
\usepackage{makecell}
\usepackage{graphicx}
\usepackage{stfloats}
\usepackage{adjustbox}

\begin{document}

\date{}

\title{\Large \bf Flashot: A Snapshot of Flash Loan Attack on DeFi Ecosystem}

%
%
\author{Yixin Cao\thanks{yixincao11@gmail.com}}
\author{Chuanwei Zou}
\author{Xianfeng Cheng}
\affil{\textit{Shanghai Wanxiang Blockchain Inc.}}


\maketitle

\begin{abstract}
Flash Loan attack can grab millions of dollars from decentralized vaults in one single transaction, drawing increasing attention from the Decentralized Finance (DeFi) players. It has also demonstrated an exciting opportunity that a huge wealth could be created by composing DeFi's building blocks and exploring the arbitrage chance. However, a fundamental framework to study the field of DeFi has not yet reached a consensus and there's a lack of standard tools or languages to help better describe, design and improve the running processes of the infant DeFi systems, which naturally makes it harder to understand the basic principles behind the complexity of Flash Loan attacks.  

In this paper, we are the first to propose \texttt{Flashot}, a prototype that is able to transparently illustrate the precise asset flows intertwined with smart contracts in a standardized diagram for each Flash Loan event. Some use cases are shown and specifically, based on \texttt{Flashot}, we study a typical \textit{Pump and Arbitrage} case and present in-depth economic explanations to the attacker's behaviors. Finally, we conclude the development trends of Flash Loan attacks and discuss the great impact on DeFi ecosystem brought by Flash Loan. We envision a brand new quantitative financial industry powered by highly efficient automatic risk and profit detection systems based on the blockchain.
\end{abstract}

\section{Introduction}
\label{sec:intro}

Flash Loan is an unprecedented outcome of Decentralized Finance (DeFi) ecosystem. The idea was first proposed with the notion \textit{Flash Lending} in July 2018 by a start-up project called \textit{Marble}~\footnote{https://medium.com/marbleorg/introducing-marble-a-smart-contract-bank-c9c438a12890}, which tried to let anyone borrow assets without collateral to take advantage of arbitrage opportunities as long as the funds are returned within the scope of the same transaction. Although some DeFi protocols may offer such function implicitly since 2019, it was first publicly implemented and made available to market by one of the popular DeFi protocols \textit{Aave}~\cite{aave}. In fact, Flash Loan did not get a lot of attention before DeFi went viral last year. Several large Flash Loan attacks accompanied the rise of the DeFi community, pushing Flash Loan to the forefront.

Before delving into how Flash Loan works and why it deserves our attention, we believe it is helpful to introduce a brief review of DeFi ecosystem first, with the hope to reach a wider audience. A more comprehensive review can be found in ~\cite{werner2021sok}.

\begin{itemize}[leftmargin=0cm,label={}]
\item \textbf{The Bright Future of DeFi}

The year 2020 saw a pure community-driven DeFi ecosystem thriving. According to DeFi Pulse~\footnote{https://defipulse.com/ (Accessed January 29$^{th}$ 2021).}, the Total Value Locked (TVL) in a sum of 63 DeFi protocols has reached {\$26.504 Billion} on January 29$^{th}$ 2021, increasing by thirty times over a year. 

DeFi refers to decentralized financial infrastructures built upon public blockchain platforms that support developing smart contracts and decentralized applications (DApps), such as \textit{Ethereum}~\cite{vitalik2013,GavinWood}. Compared with the traditional finance, in DeFi systems transactions can be settled in an atomic and transparent way thanks to the blockchain technology. Moreover, intermediaries and centralized institutions, such as custodian and central counterparty clearing house, are replaced by smart contracts that can automatically running on the blockchain platform. Built upon the innovative technical features, DeFi shows a promising potential to turn into a more open and efficient financial ecosystem with less counterparty risk~\cite{coingeckobook,schar,werner2021sok,JBVIchen}.

\item \textbf{Assets in DeFi Market}

Assets in DeFi markets are often called digital assets or virtual assets. They can be classified into two kinds, native protocol assets and tokenized assets. Bitcoin (BTC)~\cite{bitcoin} as the first successful decentralized cryptocurrency is a typical native protocol asset, which is entirely created by a blockchain protocol and gains its market value mainly from community adoption. Tokenized asset, or token for short, is on the other hand a mapping of certain asset or derivative product, to the blockchain through a process called \textit{Tokenization}. There are various ways to create tokens~\cite{tokenization1,tokenization2,xuefeng}. The most adopted tokens are stable coins. For example, USDT~\footnote{https://tether.to/wp-content/uploads/2016/06/TetherWhitePaper.pdf} and USDC~\footnote{https://www.circle.com/en/usdc} backed by off-chain fiat money share the largest market capitalization in this section~\footnote{according to coinmarketcap.com, last accessed Jan 29$^{th}$ 2021}. DAI is a decentralized stable coin backed by over-collaterized digital assets~\footnote{https://makerdao.com/en/whitepaper/}. In fact, the vast majority of tokens are issued on \textit{Ethereum}~\cite{tokenization2} through a smart contract template referred to as the ERC-20 token standard~\footnote{https://eips.ethereum.org/EIPS/eip-20}. The major purpose of tokenization is to reduce transaction frictions and make assets more accessible and flexible. This feature contributes to make DeFi a more efficient financial ecosystem. 

\item \textbf{Smart Contract}

In \textit{Ethereum}, a smart contract is intrinsically a piece of code created according to a protocol and broadcasted to the blockchain network. The code is first written in a human readable way~\footnote{\texttt{Solidity} is one of the most used programming languages for writing smart contracts on \textit{Ethereum}.} to specify the pre-agreed details such as transaction rules and execution restrictions \textit{etc}. And then it undergoes translation to a machine readable level in order to be executed in a temper-proof environment named as \textit{Ethereum Virtual Machine} (EVM) when it is called. The distributed ledger of blockchain platform also makes sure the uploaded smart contract itself is immutable unless some guy changes it using an admin key, if there's any~\footnote{Admin key is not a necessary part of smart contract. It is supposed to allow a set of predefined key holders to upgrade the DeFi protocols or perform urgent shutdowns when some governmental condition is satisfied. Usually the admin key is set with multi-sig and time-locks and controlled by project's core team. Even though the team has no intension to do malicious things, there're still potential risks from third parties.}.

Smart contract is created or called by an electronic-signed transaction sent by the creator's or user's address~\footnote{analogy to the bank account in the fiat money system,
but the entity who owns it is kept secret. A blockchain address is generated from the public-private key pair created by asymmetric cryptography. The private key is used to sign the transaction sent by the corresponding address, which authenticates that the transaction can only be sent by the private-key holder.} to the blockchain platform, while in the transaction the user specifies the operation instructions and required parameters. It can also be called by internal transactions from another running smart contract and there's no need for user to re-sign. In short, if well designed, a single electronic-signed transaction can trigger a series of smart contracts and internal transactions to be executed in a deterministic, automatic way.

If the transaction is successfully executed, it will change the states of related contract parameters and asset balances as secured in blockchain addresses will be updated accordingly. Once any of the predefined restrictions is breached during the execution, the transaction will be reverted and the states will remain the same as if this transaction hadn't taken place. Only a transaction fee will be charged from the address that sends this transaction.

\item \textbf{DeFi's Building Blocks}

DeFi is actually composed by a bunch of building blocks, or \textit{money legos}. Some of them can find prototypes in traditional Finance, while others are original, such as \textit{Uniswap}~\footnote{\textit{Uniswap} is a decentralized exchange that facilitates automatic swapping of tokens, where price is determined according to a pre-defined constant product formula and liquidity provided by different investors is homogenized in a same liquidity pool.}~\cite{uniswap}. Nevertheless, they are all financial products or services in a decentralized version, being translated to smart contracts and interacting with users through DApps~\footnote{DApps are frontend apps that interface with smart contracts through ABI.}. 

The key concept by using the term \textit{money legos} is to emphasize the composability of the DeFi protocols. \textit{Money legos} can be composed to form various kinds of systems in DeFi and work together with the common base settlement layer of the blockchain. Up to now, a few basic types of \textit{money legos} have seen promising market, \textit{e.g.}, stable coin, decentralized exchange(DEX), lending\&borrowing, derivatives, asset management, insurance \textit{etc}. They are classified to be within the application layer or protocol layer in a multi-layered framework proposed by some researchers~\footnote{https://medium.com/pov-crypto/ethereum-the-digital-finance-stack-4ba988c6c14b}~\cite{schar,werner2021sok}. The multi-layered framework gives a conceptual overview of the different constructs of DeFi ecosystem on a macroscopic scale. But one may still feel confused about how the DeFi systems really work and how Flash Loan exploiters find a vulnerability in them. There is a need of more specific descriptions probing on a microscopic scale to help people get started.

\item \textbf{Flash Loan}

Some aggregation DApps have already shown opportunities and risks the composability may bring by offering on-chain asset management services or repackaging assets to form mutilayer-structured financial instruments~\cite{crisis}. Flash Loan amplifies it in a dramatic way. In one single transaction, it can either facilitate a liquidation bot to gain \$3.6M as a revenue, or be utilized by a smart exploiter to grab millions of dollars from a complex composite of DeFi protocols. There are only a few inspiring early works~\cite{thesis,attackQin,flashWang,werner2021sok} studying on the Flash Loan attack. Wang \textit{et al.} ~\cite{flashWang} proposed a 3-phase transaction-based analysis framework to identify Flash Loan transactions by applying observed transaction patterns and tried to reveal the senders' intentions through a behavior classifier. Gronde~\cite{thesis} presented a more thorough analysis about Flash Loan's market and applications but not up to date. Qin \textit{et al.} ~\cite{attackQin} treated Flash Loan attack as an optimization problem and proposed a model formalized by a state transition function with constraints. Werner \textit{et al.} ~\cite{werner2021sok} presented a systematic introduction of attacks in DeFi ecosystem, including those exploiting Flash Loan. In this paper, we hope to shed light on some more aspects. 
\end{itemize}

This paper mainly contributes in the following aspects,
\begin{itemize}
\item We propose a prototype called \texttt{Flashot} to uncover the microscopic process of the Flash Loan attack in a clear and precise way. In fact, \texttt{Flashot}  is created to be a standard tool that can be utilized to illustrate asset flows in any kind of DeFi systems.
\item An in-depth analysis about a typical Flash Loan attack targeted at \textit{bZx}  is presented based on \texttt{Flashot}, with a more accurate model and solution regarding the optimization problem plus some economic explanations to the attacker's behaviors.
\item We summarize the development trends of Flash Loan attacks and discuss about some risk control strategies.
\item Finally, we envision a groundbreaking financial ecosystem where Flash Loan plays a significant role.
\end{itemize}

The rest of the paper is arranged as follows. In Section~\ref{sec:background}, we first give a background about what Flash Loan is and compare its properties to other products in debt market. A list of Flash Loan attack events that have occurred is also collected in Section~\ref{subsec:events}.  In Section~\ref{sec:flashot}, we propose a novel prototype called \texttt{Flashot} to draw the running processes of Flash Loan transaction in a single diagram. A typical case is studied based on \texttt{Flashot} and presented in Section~\ref{sec:casestudy} with an in-depth analysis in Section~\ref{subsec:economics}. In Section~\ref{sec:trends}, we show the trends of Flash Loan attacks with \textit{flashots} of eight additional cases attached in appendix~\ref{appendix}. Finally, we make a conclusion and share some discussions in Section~\ref{sec:conclusion}.

\section{Background}
\label{sec:background}

Currently, Flash Loan services are provided by four representative DeFi protocols namely \textit{Aave}~\footnote{https://docs.aave.com/developers/guides/flash-loans}, \textit{dYdX}~\footnote{https://help.dydx.exchange/en/articles/3724602-flash-loans}, \textit{Uniswap V2}\footnote{https://uniswap.org/docs/v2/core-concepts/flash-swaps/}, and \textit{bZx}~\footnote{https://github.com/bZxNetwork/flashloan-sample}. To use Flash Loan, one need to send a transaction calling a smart contract to conduct all of the operations. The basic logic includes three steps, \textit{i.e.}
\begin{enumerate}[leftmargin=*,label=(\roman*)]
\item Lend flash loan without collateral or credit certification; 
\item Make use of the flash loan to gain a profit; 
\item Repay flash loan plus interest.
\end{enumerate}

As discussed about smart contract in Section~\ref{sec:intro}, if in step (ii) one fails to make extra money, he will not have enough money to fulfill step (iii), so that the transaction will be reverted~\footnote{Unless the user pays to fill the difference, for example, in a wash trading.}. The most important thing to be noted here is that the borrowing of flash loan is not valid unless the flash loan can be fully repaid within the scope of the same transaction, which technically eliminates the default risk. 

There's no Flash Loan equivalent in the real world. Default risk exists in any loan products in the traditional financial market and it raises the interest rate to different extent. Decentralized lending products provided by DeFi protocols such as \textit{Maker} and \textit{Compound}~\footnote{https://compound.finance/} prevent it by requiring over-collaterization of digital assets, nevertheless users still have to encounter losses when the collaterized assets are not liquidated as soon as possible, especially when the market fluctuates sharply. Flash Loan, on the contrary, is a business that never loses money.

\subsection{Properties of Flash Loan}

Given the technical advantage, Flash Loan offers a rather convenient borrowing service with very low interest rate~\footnote{0.09\% for \textit{Aave}, 0.3\% for \textit{Uniswap V2}, 2 Wei for \textit{dYdX}, and free for \textit{bZx}.}. Moreover, users can borrow as much as the total amount available in flash loan pools. There's no limitation on the borrowing amount as long as you can repay back. Flash Loan is also accessible easily by anyone without complicated censorship.

In a word, Flash Loan can provide abundant of assets at a very low interest rate in flash speed to anyone. These properties make Flash Loan an ideal source of funds for attackers, and the trial and error cost of the attack is very low, just some transaction fee charged by the base settlement layer of the blockchain platform.

\subsection{Flash Loan Events}
\label{subsec:events}

As listed in Table~\ref{table:events}, several influential events exploiting Flash Loans have occurred since February 2020. Qin \textit{et al.} ~\cite{attackQin} provided a detailed analysis on two attacks happened in February 2020 and categorized them as \textit{Pump and Arbitrage} attack and \textit{Oracle Manipulation} attack respectively. We follow the notions in classifying subsequent cases that occurred and find that most of them correspond to a manipulation of oracles. Two events were associated with a reentrancy attack~\cite{tubiblio111410}. The exact DeFi protocols implicated in these events and the estimated value of proceeds grabbed by attackers at the time are shown as well. From the data, a more than tenfold increase in the profit of a single attack is witnessed.

\begin{center}
\begin{table*}[h]
\caption{Striking events exploiting Flash Loan}
\label{table:events}
\centering
\begin{tabularx}{1\textwidth}{l|l|l|l|l|l}

\hline\hline
\makecell[c]{Date}	&	\makecell[c]{Event Label}   & 		\makecell[c]{Flash Loan Provider}		& 	\makecell[c]{Implicated DeFi Protocols} &	\makecell[c]{Type}	& 	\makecell[c]{Proceeds}\\
\hline
\multirow{4}{*}{2020-02-15}	&	\multirow{4}{*}{bZx Pump Attack}	&	\multirow{4}{*}{\textit{dYdX}}	&	\textit{bZx}			&	\multirow{4}{*}{Pump and Arbitrage}	&	\multirow{4}{*}{\$330K}\\
						&							&								&	\textit{Compound}		&								&					\\
						&							&								&	\textit{Kyber}			&								&					\\
						&							&								&	\textit{Uniswap V1}		&								&					\\
\hline
\multirow{4}{*}{2020-02-18}	&	\multirow{4}{*}{bZx Oracle Attack}	&	\multirow{4}{*}{\textit{bZx}}		&	\textit{bZx}			&	\multirow{4}{*}{Oracle Manipulation}	&	\multirow{4}{*}{\$638K}\\
						&							&								&	\textit{Uniswap V1}		&								&					\\
						&							&								&	\textit{Kyber}			&								&					\\
						&							&								&	\textit{Synthetix}		&								&					\\
\hline
\multirow{2}{*}{2020-06-28}	&	\multirow{2}{*}{Balancer Attack} &	\multirow{2}{*}{\textit{dYdX}}		&	\textit{Balancer}			&	\multirow{2}{*}{Pump and Arbitrage}	&	\multirow{2}{*}{\$439K}\\
						&							&								&	\textit{Uniswap V2}		&								&					\\
\hline
\multirow{3}{*}{2020-10-26}	&	\multirow{3}{*}{Harvest Attack}	&	\multirow{3}{*}{\textit{Uniswap V2}}	&	\textit{Harvest}			&	\multirow{3}{*}{Oracle Manipulation}	&	\multirow{3}{*}{\$26.6M}\\
						&							&								&	\textit{Curve}			&								&					\\
						&							&								&	\textit{Uniswap V2}		&								&					\\
\hline
\multirow{2}{*}{2020-11-06}	&	\multirow{2}{*}{Cheese Bank Attack}&	\multirow{2}{*}{\textit{dYdX}}	&	\textit{Cheese Bank}		&	\multirow{2}{*}{Oracle Manipulation}	&	\multirow{2}{*}{\$3.3M}\\
						&							&								&	\textit{Uniswap V2}		&								&					\\
\hline
\multirow{1}{*}{2020-11-12}	&	\multirow{1}{*}{Akropolis Attack}&	\multirow{1}{*}{\textit{dYdX}}		&	\textit{Akropolis}		&	\multirow{1}{*}{Reentrancy}		&	\multirow{1}{*}{\$2M}\\

\hline
\multirow{4}{*}{2020-11-14}	&	\multirow{4}{*}{Value.DeFi Attack}&								&	\textit{Value.DeFi}		&	\multirow{4}{*}{Oracle Manipulation}	&	\multirow{4}{*}{\$7.4M}\\
						&							&	\textit{Aave}					&	\textit{Curve}			&								&					\\
						&							&	\textit{Uniswap V2}				&	\textit{Uniswap V2}		&								&					\\
						&							&								&	\textit{SushiSwap}		&								&					\\
\hline
\makecell[l]{2020-11-17}		&	\makecell[l]{OUSD Attack}		&	\makecell[l]{\textit{dYdX}}			&	\makecell[l]{\textit{OUSD}\\\textit{Uniswap V2}\\\textit{SushiSwap}}	&	\makecell[l]{Reentrancy}	&\makecell[l]{\$7.9M}\\

\hline
\makecell[l]{2020-12-18}		&\makecell[l]{Warp Finance Attack}	&\makecell[l]{\textit{Uniswap V2}\\\textit{dYdX}}&\makecell[l]{\textit{Warp Finance}\\\textit{Uniswap V2}\\\textit{Sushiswap}}		&	\makecell[l]{Oracle Manipulation}	&	\makecell[l]{\$941K}\\

\hline\hline

\end{tabularx}
\end{table*}
\end{center}

\section{Flashot and Example}
\label{sec:flashot}
Actually, even the senior DeFi players may find it difficult to understand what really happened during the Flash Loan attack events as described in Section~\ref{subsec:events}. The DeFi ecosystem is developing too fast to wait for the creation of useful research tools. People are still trying to understand and discuss the phenomena emerged in this field in rather native ways. Unlike industries that have developed for centuries, there are no applied tools fit for the nascent DeFi ecosystem, such as a circuit diagram for electronic engineering or an engineering drawing for mechanical engineering. 

Decentralized computing platforms, \textit{Ethereum} as a pioneer, provide a promising opportunity to transform the traditional financial systems that heavily relying on manual works into a financial engineering industry. One of the core components is smart contract. Smart contracts interact with each other and automatically execute and record the transactions between assets. It will be useful and foresightful to form a universal language describing the functions and events of smart contracts, and to outline where assets are flowing and how they are transformed in each transaction.
  
To our best knowledge, we are the first to propose a standardized asset flow diagram called \texttt{Flashot}, which is designed to reveal the asset flows among various protocols in a transparent and precise way. For a comparison, similar attempts in  ~\cite{flashWang}(Fig. 4 and Fig. 5 therein) and ~\cite{attackQin} (Fig. 6 and Fig. 7 therein) are either too simple to uncover important messages or hard to understand and standardize.

\subsection{Flashot}
\texttt{Flashot} is composed of three basic elements and two classes of operations extracted from the DeFi systems. 

\begin{itemize}[leftmargin=0cm,label={}]
\item \textbf{Elements:}

\begin{enumerate}[leftmargin=*,label=(\arabic*)]
\item \textbf{Asset} corresponds to the a naitve protocol asset created on the blockchain such as ETH or a tokenized asset such as the stable coin USDC. Each kind of asset has a ticker to distinguish with each other.
\item \textbf{Smart Contract} corresponds to a piece of code that automatically executed on the decentralized blockchain platform, according to certain protocol. Each smart contract has a name and a unique Hash as its index in \textit{Ethereum}. Here for legibility, we label smart contracts by their names as an example.
\item \textbf{Asset Pool} corresponds to a blockchain address that records assets' balances, functioning as a vault. Asset Pools include various types, \textit{e.g.}, Lending Pools, Liquidity Pools, Minting Pools. There may be one or several kinds of assets in an asset pool.
\end{enumerate}

\item \textbf{Operations:}
\begin{enumerate}[leftmargin=*,label=(\arabic*)]
\item \textbf{Split/Merge} Assets can be either splited into parts or merged into a bulk.
\item \textbf{Transform} One kind of asset can be transformed into another by calling one or more smart contracts. Transform operations include swapping, depositing, redeeming, withdrawing, \textit{etc}.
\end{enumerate}

\end{itemize}

\begin{figure*}
\centering
\includegraphics[width=1\textwidth]{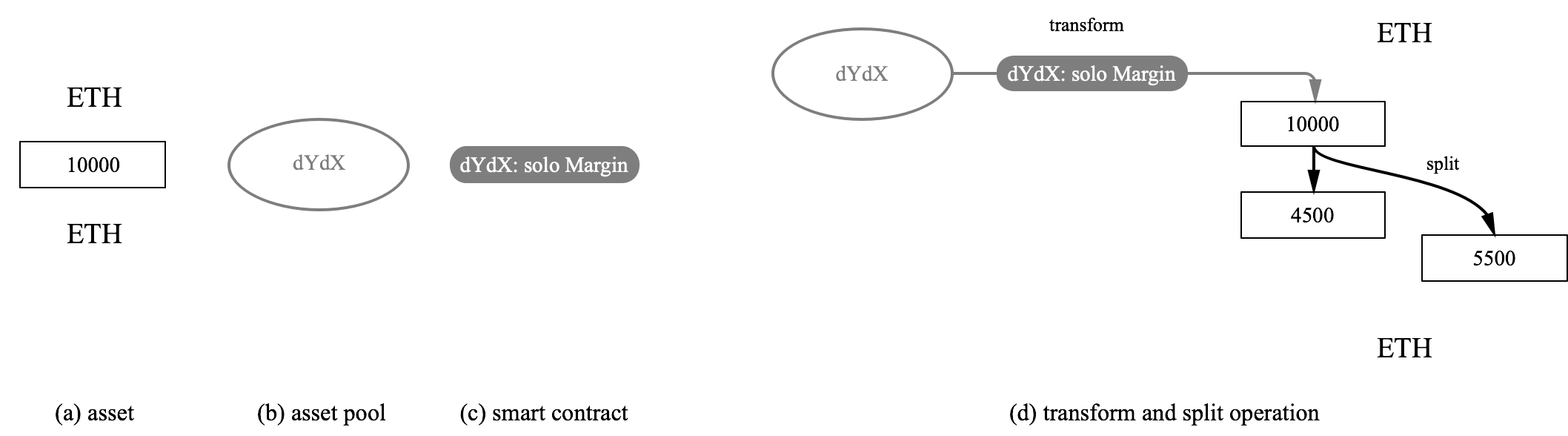}
\caption{\label{fig:operation} (a) Asset. An amount of 10,000 ETH is shown in this example. (b) Asset pool. The protocol \textit{dYdX}'s asset pool is shown in this example. (c) Smart contract. In this example, the contract name is \textit{dYdX: solo Margin}. (d) Transform operation is denoted by a horizontal arrow line connecting an asset/asset pool to another, with smart contract's name attached. A split operation is denoted by vertical arrow lines.}
\end{figure*}

As shown in Fig.~\ref{fig:operation}(a)$-$(c), we use a rectangle to represent a sum of asset and the figures inside it represent the amount. To distinguish different kinds of assets, the rectangles will also be labeled with the corresponding tickers. An oval shape represents an asset pool with the protocol's name inside it. Smart contract is represented by a rounded rectangle. 

Spit or merge operations are denoted by vertical arrow lines connecting rectangles representing a same kind of assets. Transform operations are denoted by horizontal arrow lines connecting either asset pools or assets, with associated smart contracts' name attached. Fig.~\ref{fig:operation}(d) shows an example.

\subsection{Example}
\begin{figure*}[h]
\centering
\includegraphics[width=1\textwidth]{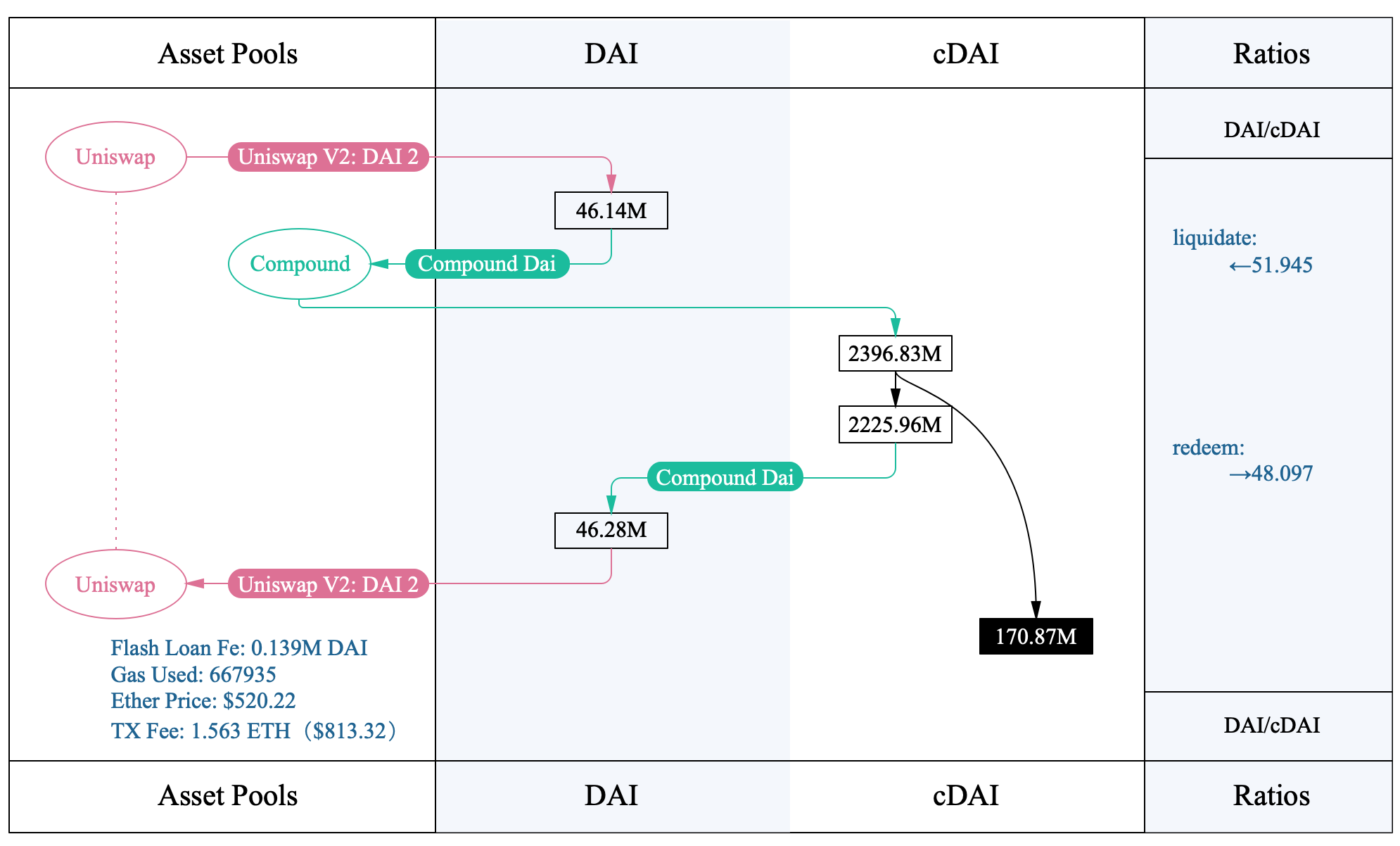}
\caption{\label{fig:compound}A \textit{flashot} of a transaction triggered by a Compound liquidation bot on November 26$^{th}$ 2020. The left panel shows asset pools that provide flash loan or interact with asset flows. The middle panel shows asset flows with associated assets arranged in each column in order of appearance. In the right panel, critical ratios are listed as a reference. Asset's split or merge operation is denoted by a set of solid arrow lines in the vertical direction. Transform operation is denoted by an arrow line with smart contract's name attached in the horizontal direction. Figures in black solid rectangle represent the amount of ultimate proceed.}
\end{figure*}

Next we use \texttt{Flashot} to illustrate a notable transaction~\footnote{txhash: 0x53e09adb77d1e3ea593c933a85bd4472371e03da12e3fec853b\\*5bc7fac50f3e4} sent by a liquidation bot during the Compound Liquidation Event on November 26$^{th}$ 2020~\footnote{https://beincrypto.com/100m-liquidated-from-compound-following-flash-loan-exploit/}. As shown in Fig.~\ref{fig:compound}, the whole process of the liquidation transaction is clearly presented. There are three panels in the \textit{flashot}, as described in the following.

\begin{itemize}
\item \textbf{Asset Pool Panel}

The left panel in the \textit{flashot} shows the asset pools that either provide flash loan or interact with asset flows if there is any. Some asset pools may be interacted more than once at different stages of the transaction process. For example, flash loan pools  are always associated with an asset flowing out and an asset flowing in. We use a dotted line to connect the same asset pool that appears at different stages.

\item \textbf{Asset Panel} 

The middle panel is the major panel illustrating asset flows. All associated assets are arranged in each column in order of appearance. As introduced earlier, asset flows are triggered by two classes of operations. Asset's split or merge operation is denoted by a set of solid arrow lines in the vertical direction. Transform operation is also denoted by an arrow line in the horizontal direction, and the rounded rectangle denoting smart contract is attached. The arrow directions point to the destination of the asset flows. Figures in the black solid rectangle represent the amount of ultimate proceed.

\item \textbf{Ratio Panel} 

In the right panel, some critical ratios or parameters will be listed as a reference. 

\end{itemize}

From Fig.~\ref{fig:compound}, we can easily find that this liquidation bot borrowed around 46 million DAI from \textit{Uniswap V2}'s DAI pool, transfered it to \textit{Compound}'s asset pool to liquidate an underwater account at a favorable liquidation ratio (1 DAI for 51.945 cDAI). Next, it splitted its asset cDAI into two parts. One part of cDAI was used to redeem DAI at a market redemption ratio(1 DAI for 48.097 cDAI). The amount of this part is carefully calculated to get just enough DAI to repay the flash loan plus an interest fee (0.3\%). The remain part of cDAI is the revenue of this liquidation bot, which equals about \$3.6M at then market price. In this diagram, the exact asset flow path from the very beginning to the end is revealed along the arrow lines.

\section{Case Study: Pump and Arbitrage }
\label{sec:casestudy}
\begin{figure*}
\centering
\includegraphics[width=1\textwidth]{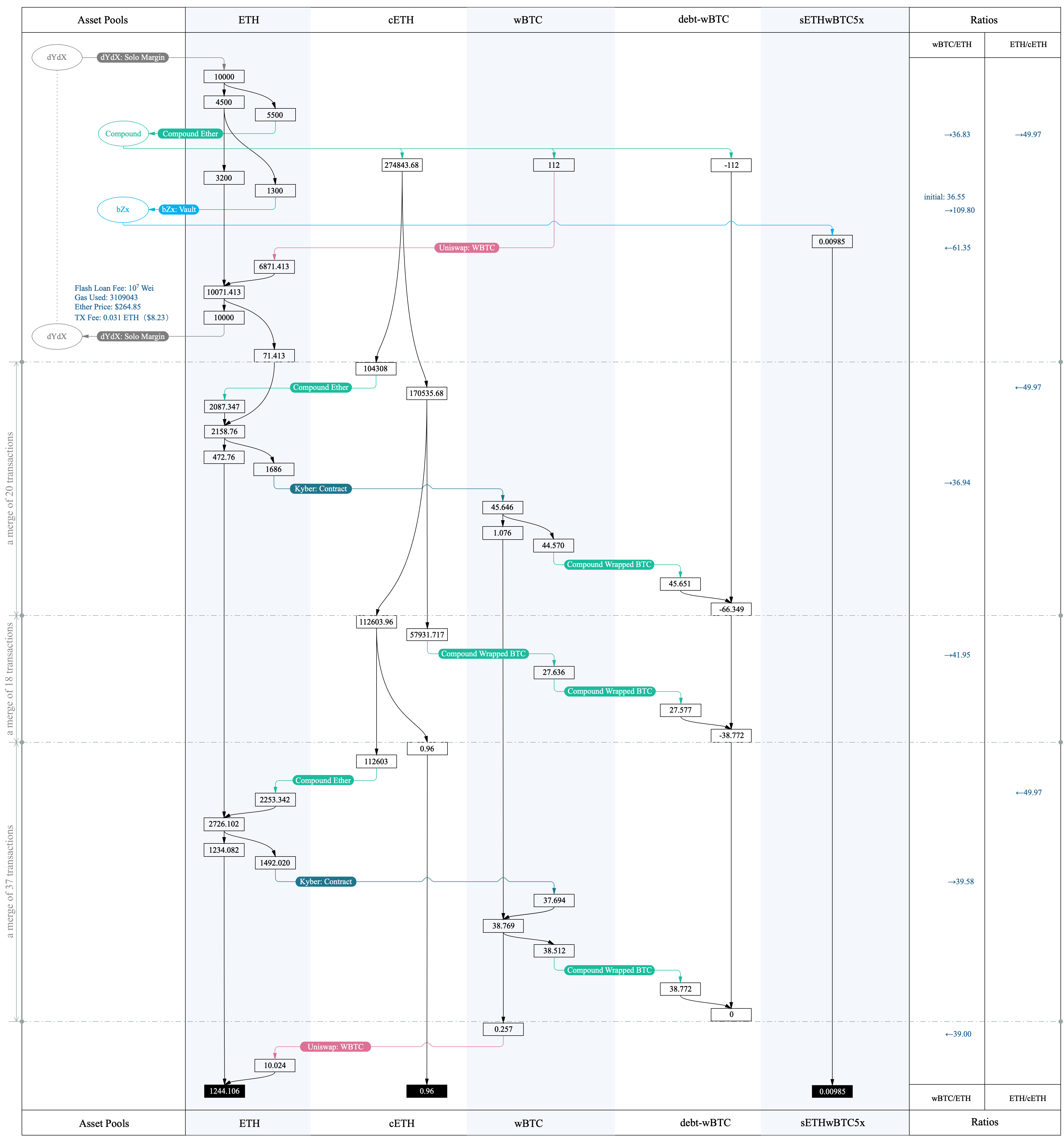}
\caption{\label{fig:bzxpump}\textit{Flashot} of bZx Pump Attack. There are five panels from top to bottom. The first panel shows the first Flash Loan transaction at block height 9484688 in which the major attack processes occurred. The second and fourth panels correspond to a total of 57 transactions sent by the attacker to redeem collateral in \textit{Compound} during block height 9484917 to 9496529. The third panel presents a merge of 18 transactions triggered by liquidation bots. The last panel shows the last swap transaction and the ultimate proceeds left in the attacker's address.}
\end{figure*}

\texttt{Flashot} is able to illustrate even more complicate process. As listed in Table~\ref{table:events}, the first well-known Flash Loan attack, bZx Pump Attack, took place on February 15$^{th}$ 2020~\footnote{First txhash: 0xb5c8bd9430b6cc87a0e2fe110ece6bf527fa4f170a4bc8cd0\\*32f768fc5219838}, which is a typical case covered by previous works~\cite{thesis,attackQin,flashWang}. Here we continue to show the power of \texttt{Flashot} based on this case, and give a guide on how to use \texttt{Flashot} for analysis.

\subsection{The Attack Process}

The \textit{flashot} corresponding to the event bZx Pump Attack is shown in Fig.~\ref{fig:bzxpump}. At the very beginning, the attacker borrowed a total of 10,000 ETH from \textit{dYdX}'s flash loan pool and splitted it into three parts. The first part, 5,500 ETH, was deposited to \textit{Compound}'s lending pool as a collateral in order to borrow 112 wBTC. At the same time, the attacker got cDAI to be used to redeem the collateral later. We can also describe this operation literally as (ETH:5,500) $\xrightarrow[]{Compound Ether}$ (cETH:274,843.68, wBTC:112, debt-wBTC:$-$112) for example. The actual exchange ratio between wBTC and ETH is 36.83 while the Collateral Factor~\footnote{In \textit{Compound}, a Collateral Factor is set for each token, the reciprocal of which indicates the over-collateral ratio.} equals 0.75.

\begin{figure*}[ht]
\centering
\includegraphics[width=1\textwidth]{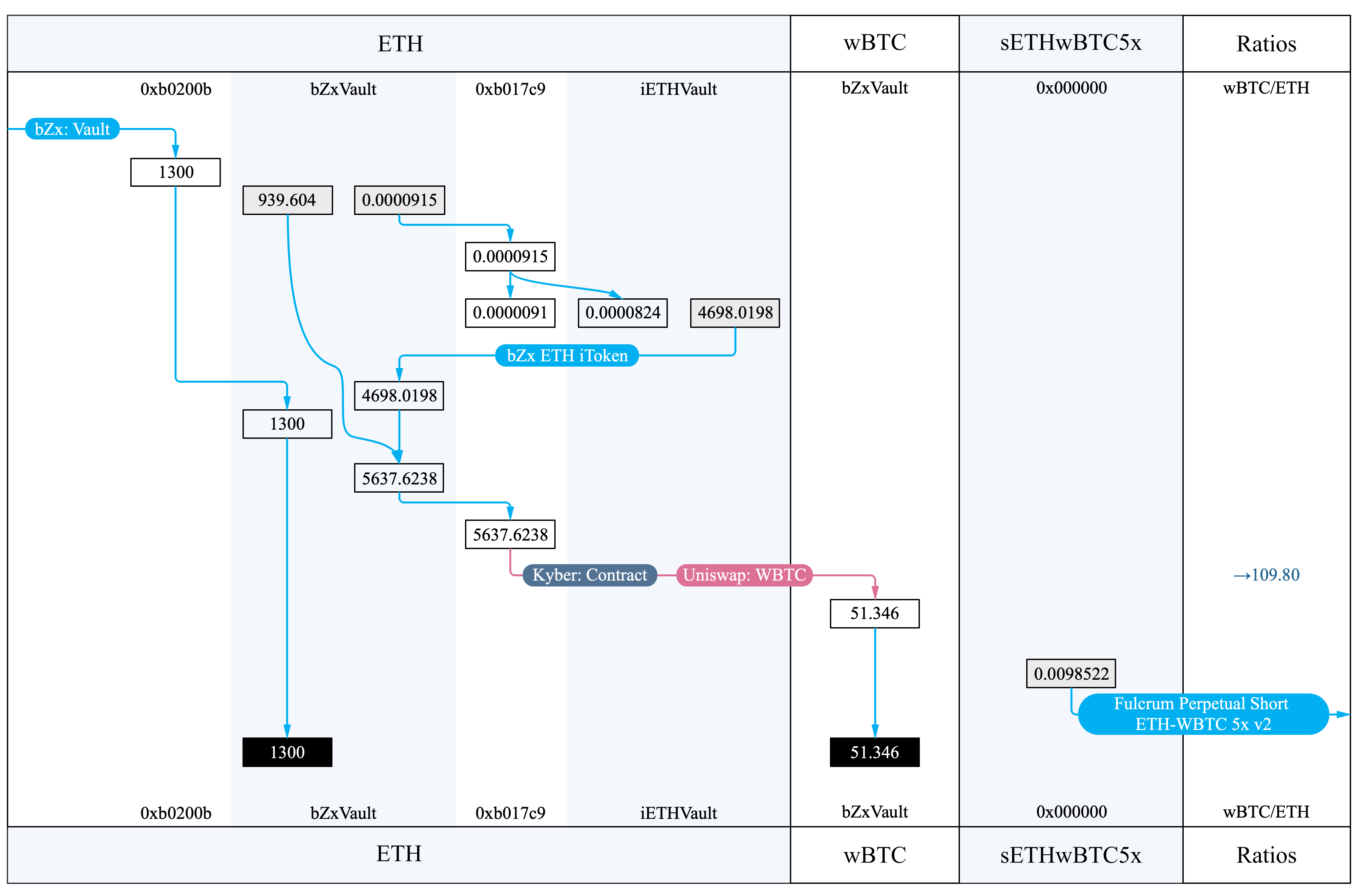}
\caption{\label{fig:bzxpumpinset}\textit{Flashot} of internal process of 5x short margin trade in \textit{bZx} protocol as exploited by the attacker in the event bZx Pump Attack.}
\end{figure*}

Then the attacker deposited the second part of flash loan, a total of 1,300 ETH, to \textit{bZx}'s vault as a margin collateral in order to short ETH in favor of wBTC at 5x leverage by calling \textit{bZx}'s smart contract. The internal process of the margin trade was handled automatically by \textit{bZx} protocol as shown in Fig.~\ref{fig:bzxpumpinset}. To execute the order, \textit{bZx}'s smart contract borrowed about 4,698.02 ETH from its iETH Vault and swapped about 5,637.62 ETH to 51.346 wBTC at the DEX \textit{Uniswap} through a router~\footnote{A router can aggregate different DEXs to find a trading path at a best price.} called \textit{Kyber}. Since \textit{Uniswap} is an automated market maker (AMM) where price is set according to a constant product formula relying on no external information, the price deviation from the initial price (36.55 in this case) will increase sharply with the trading volume in a swap transaction, giving the attacker a chance to manipulate the market. By this margin trade, he pumped the average price of wBTC at \textit{Uniswap} by three times up to 109.8 ETH and get some derivative-like token called sETHwBTC5x~\footnote{contract address: 0xb0200b0677dd825bb32b93d055ebb9dc3521db9d}.

To take advantage of the price spread between \textit{Uniswap} and other market, 112 wBTC borrowed from \textit{Compound} was immediately swapped back to 6871.413 ETH at a ratio of 61.35, which is about 1.67 times of 36.83. The attacker merged 6,871.413 ETH with the remain third part, 3,200 ETH and repaid 10,000 ETH plus a flash loan fee ($10^{7}$ Wei).

As a result of this transaction, the attacker gained 71.413 ETH directly. Meanwhile, he owned 274,843.68 cETH, which can be used to redeem the collaterized ETH in \textit{Compound} after paying back the borrowed 112 wBTC plus interest. He also had 1,300 ETH deposited in \textit{bZx}'s vault as a margin and a position of 51.346 wBTC in the margin trade. He could close the position by burning sETHwBTC5x in his address as long as it were not liquidated.
 
Actually, 75 transactions were executed in several subsequent blocks to pay back 112 wBTC and redeem ETH. 57 of them were actively sent by the attacker, which went through a three-step process, cETH $\xrightarrow[]{Compound Ether}$ ETH $\xrightarrow[]{Kyber: Contract}$ wBTC $\xrightarrow[]{Compound Wrapped BTC}$ debt-wBTC. It is also interesting to see that there were 18 transactions corresponding to a liquidation process executed by third parties. As for the position and margin collateral in \textit{bZx}, the attacker just left them alone.

Finally, the proceeds in this Flash Loan attack event was about 1244.106 ETH (about \$ 330K at then market price), neglecting a small amount of cETH and sETHwBTC5x.

\subsection{Economics Behind Behaviors}
\label{subsec:economics}
The essential reason why Flash Loan attack can succeed is because DeFi's building blocks are automatically running based on predefined algorithms and predictable parameters. It's capable to fomulate the problem as an optimization equation under a few constraints. Qin \textit{et al.} ~\cite{attackQin} had already shared an instructive analytical framework in this aspect, despite the flaw that they analyzed the optimization problem based on procedures and parameters over several blocks, which introduced future parameters in the calculation process.

In this paper, we offer a more accurate version of derivation as well as economic explanations behind the attacker's behaviors.

According to the \textit{flashots} in Fig.~\ref{fig:bzxpump} and Fig.~\ref{fig:bzxpumpinset}, we break down the first transaction of bZx Pump Attack into 6 steps, \textit{i.e.},

\begin{enumerate}[leftmargin=*,label=(\arabic*)]

\item Borrow $n$ ETH from \textit{dYdX};

\item Deposit $n_{1}$ ETH as collateral to borrow $\frac{cf\cdot{n_{1}}}{p}$ wBTC from \textit{Compound}, where $0<cf\leq 1$ is the Collateral Factor set by \textit{Compound}~\footnote{https://compound.finance/markets/ETH}, $p$ is the market price of wBTC;

\item Borrow $(\frac{l}{ocr}-1)(n-n_{1})$ ETH from \textit{bZx} by depositing $(n-n_{1})$ ETH as margin collateral~\footnote{Here we suppose the flash loan is used up, which is the optimal choice.}, where $ocr>1$ denotes the Over Collateral Ratio set by \textit{bZx};

\item Short sell $\frac{l \cdot (n-n_{1})}{ocr}$ ETH and get $\Delta b$ wBTC through \textit{Uniswap};

\item Swap $\frac{cf\cdot{n_{1}}}{p}$ wBTC to $\Delta e$ ETH through \textit{Uniswap};

\item Repay the Flash Loan.
\end{enumerate}

After setp (6), the attacker obtained $P_{f}=\Delta e-n$ ETH as a direct proceed, an amount of cETH and some sETHwBTC5x. He may choose another two steps to close the margin trade in \textit{bZx} and repay the debt in \textit{Compound}, 

\begin{enumerate}[leftmargin=*,label=(\arabic*)]
\setcounter{enumi}{6}
\item Close the short position by selling $\Delta b$ wBTC in exchange for ETH. Repay $(\frac{l}{ocr}-1)(n-n_{1})$ ETH plus interest. The remain ETH deducted by $(n-n_{1})$ ETH will be net profit of the margin trade, if there is any;

\item Repay $\frac{cf\cdot{n_{1}}}{p}$ wBTC to redeem $n_{1}$ ETH.

\end{enumerate}

The key to success is to finish steps (1)$-$(6) in one transaction without revertion. Steps (7)$-$(8) are actually like options offered to the attacker, who can decide whether to exercise them in a favorable way. While Qin \textit{et al.} ~\cite{attackQin} included steps (1)$-$(7) to find the overall optimal output, we treat steps (1)$-$(6) as an independent optimization problem discussed in Section~\ref{subsub:optm1} and Section~\ref{subsub:optm2}. Step (7) and step (8) are treated as two additional problems related to contingent choices, which will be discussed in Section~\ref{subsub:contingent}.

\subsubsection{The Optimization Problem}
\label{subsub:optm1}
Let's suppose there are $b$ wBTC and $e$ ETH in \textit{Uniswap}'s asset pool before the Flash Loan attack takes place, and these two parameters obey the constant product formula~\footnote{https://hackmd.io/@HaydenAdams/HJ9jLsfTz} as shown in Eq.~\ref{eq1}. 

\begin{equation}
\label{eq1}
b \cdot e = k
\end{equation}

Neglecting any fee charged, after step (4), the constant product formula becomes Eq.~\ref{eq2}, where the price of wBTC is pumped from $\frac{e}{b}$ ETH to $\frac{e+\frac{l\cdot (n-n_{1})}{ocr}}{b-\Delta b}$ ETH so that a profit can be made in step (5). 

\begin{equation}
\label{eq2}
\left(b-\Delta b\right)\cdot\left[e+\frac{l\cdot (n-n_{1})}{ocr}\right]=k
\end{equation}

From Eq.~\ref{eq1} and Eq.~\ref{eq2}, we get Eq.~\ref{eq3}, 

\begin{equation}
\label{eq3}
\Delta b = k\cdot\frac{\frac{l\cdot (n-n_{1})}{ocr}}{e\left[e+\frac{l\cdot (n-n_{1})}{ocr}\right]}
\end{equation}

After step (5), the constant product formula becomes Eq.~\ref{eq4}. 

\begin{equation}
\label{eq4}
\left(b-\Delta b + \frac{cf\cdot n_{1}}{p} \right)\cdot \left[ e + \frac{l\cdot (n-n_{1})}{ocr} - \Delta e \right] = k
\end{equation}

Combining Eq.~\ref{eq2} $-$ Eq.~\ref{eq4}, we get Eq.~\ref{eq5}.

\begin{equation}
\label{eq5}
\Delta e = \frac{\frac{cf\cdot n_{1}}{p} \left[e + \frac{l\cdot (n-n_{1})}{ocr} \right]^{2}}{k + \frac{cf\cdot n_{1}}{p}\left[e + \frac{l\cdot (n-n_{1})}{ocr}\right]}
\end{equation}

In step (6), it is required that there must be enough ETH to repay the flash loan plus an interest. Thus we get the first constraint as shown in Eq.~\ref{eq6},

\begin{equation}
\label{eq6}
\Delta e > n
\end{equation}
which can be transformed to Eq.~\ref{eq7} by substituting $\Delta e$ according to Eq.~\ref{eq5}. 

\begin{multline}
\label{eq7}
\frac{cf\cdot l^2}{{ocr}^2}{(n-n_{1})}^2 + \frac{cf\cdot l}{ocr}(2e - n)(n-n_{1}) + cf\cdot e\left(e-n\right)\\
> \frac{p\cdot k\cdot n}{n_{1}}
\end{multline}

\begin{center}
\begin{table*}[h]
\caption{Parameters in bZx Pump Attack}
\label{table:param}
\centering
\begin{tabularx}{0.65\textwidth}{c|r|l}

\hline\hline
\makecell[c]{Parameter}			&	\makecell[c]{Value}   			& 		\makecell[c]{Description}\\
\hline
$cf$							&	0.75						&		Collateral Factor set by \textit{Compound}\\
$ocr$						&	1.153					&		Over Collateral Ratio set by \textit{bZx}\\
$l$							&	5						&		Leverage of margin trade in \textit{bZx}\\
$b$							&	77.08					&		Initial balance of wBTC in \textit{Uniswap}\\
$e$							&	2,817.77					&		Initial balance of ETH in \textit{Uniswap}\\
$p$							&	36.48					&		Collateralized borrowing exchange rate in \textit{Compound}\\
$p_{m}$						&	39.08					&		Market price of WBTC after the attack\\
$n_{f}$						&   10,000						&		Maximum amount of ETH to flash loan\\
$n_{c}$						&	155.7					&		Maximum amount of wBTC in \textit{Compound} to borrow\\
$n_{b}$						&   4,858.74					&		Maximum amount of ETH to leverage\\

\hline\hline

\end{tabularx}
\end{table*}
\end{center}

Eq.~\ref{eq7} gives a necessary but not sufficient condition to make sure the Flash Loan transaction will not be reverted. There are a few more constraints to follow as listed in Eq.~\ref{eq8}. These constraints restrict that the amounts of borrowed assets are positive numbers and do not exceed the available balances of the corresponding asset pools.

\begin{equation}
\label{eq8}
\begin{cases}
n \leq n_{f} \\
0< n_{1} < n\\
\frac{cf\cdot{n_{1}}}{p} \leq n_{c}\\
(\frac{l}{ocr}-1)(n-n_{1})\leq n_{b}
\end{cases}
\end{equation}

Parameters $cf$, $ocr$, $l$ are known constants set by \textit{Compound} and \textit{bZx}. $b$, $e$, $p$, $n_f$, $n_c$ and $n_b$ are predictable variables which can be fetched from the \textit{Ethereum}'s event logs. $k$ can be calculated according to Eq.~\ref{eq1}. These parameters are collected from ~\cite{attackQin} and listed in Table~\ref{table:param}. One thing to be noted is that $p$ is slightly different from the actual value (as shown in Fig.~\ref{fig:bzxpump}) calculated according to transaction logs. Since the difference will not cause a qualitative change in the final result, we will explore it in later works.

The constraints in Eq.~\ref{eq8} can be degenerated to be Eq.~\ref{eq9}, and then Eq.~\ref{eq10} after substituting the values.

\begin{equation}
\label{eq9}
\begin{cases}
n \leq n_{f}\\
\max{[0,n-\frac{n_b}{\frac{l}{ocr}-1}]} \leq n_{1} \leq \min{[\frac{n_{c}\cdot p}{cf},n]}\\
n-\frac{n_b}{\frac{l}{ocr}-1} \leq \frac{n_{c}\cdot p}{cf}
\end{cases}
\end{equation}

\begin{equation}
\label{eq10}
\begin{cases}
n - 1456.23 \leq n_{1} \leq 7573.25  &  7573.25 < n \leq 9029.48 \\
n - 1456.23 \leq n_{1} \leq n & 1456.23 < n \leq 7573.25\\
0 \leq n_{1} \leq n & 0 < n \leq 1456.23
\end{cases}
\end{equation}

Under these constraints, parameters can be tuned to reach the optimal output. Here we define the ultimate gross profit $P_{g}$ to be composed by four parts as expressed in Eq.~\ref{eq11},

\begin{equation}
\label{eq11}
P_{g}= P_{f} + n_{1}\cdot (1-cf) + P_{c1} + P_{c2}
\end{equation}
where $P_{f}=\Delta e-n$ is the direct proceed of the Flash Loan transaction as defined before, and $n_{1}\cdot (1-cf)$ is a fixed profit in the form of collateral in \textit{Compound} under the assumption that the redemption ratio is equal to the borrowing exchange rate. $P_{c1}$ and $P_{c2}$ are unpredictable extra parts associated with the contingent choices offered in step(7) and step(8) accordingly and we discuss about it in Section~\ref{subsub:contingent}.

Therefore, the main optimization problem is to maximize the predictable profit $P_{p} = P_{f} + n_{1}\cdot (1-cf)$ resulted from step (1)$-$(6).

\subsubsection{Practice of Optimization}
\label{subsub:optm2}

\begin{figure}[ht]
\centering
\includegraphics[width=1\columnwidth]{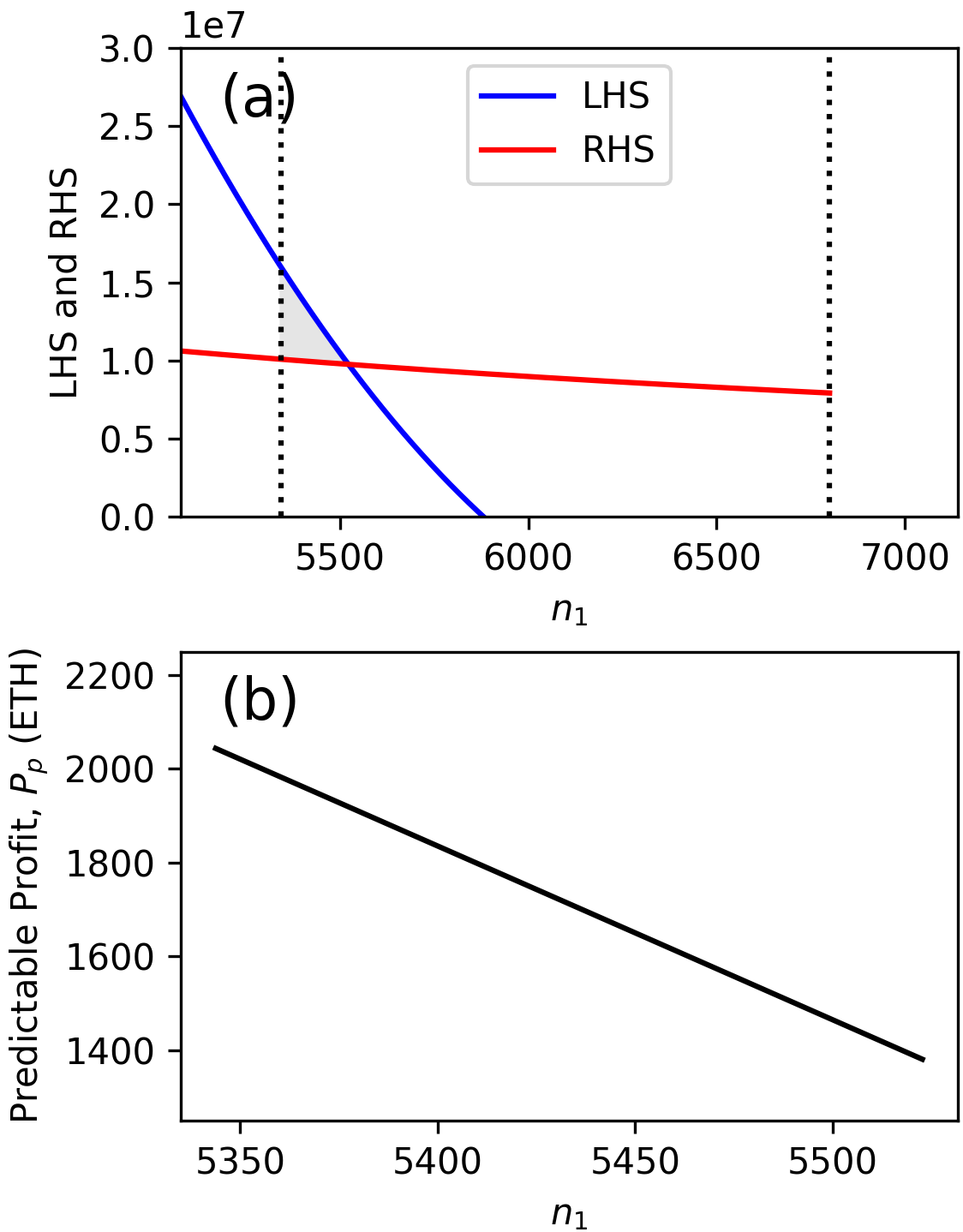}
\caption{\label{fig:condition}(a) Condition curves for a successful Flash Loan attack when $n=6,800$. The blue and red solid lines denote the LHS and RHS of Eq.~\ref{eq7}. Dotted lines denote the limitations of  $n_{1}$ according to Eq.~\ref{eq10}. The gray area shows the favorable interval of $n_{1}$ where the Flash Loan transaction will not be reverted. We see that the parameter chose by the attacker in the real event, $n_{1}=5,500$ lies in this interval. (b) Predictable profit $P_p$ versus $n_{1}$ given $n=6,800$.}
\end{figure}

Next, we show why the parameters chosen by the attacker is not optimal. In this case, $n_{1}=5,500$, $n=5,500+1,300=6,800$. It is obvious that the attacker hadn't used up the borrowed flash loan. Moreover, as expressed by Eq.~\ref{eq7} and Eq.~\ref{eq10}, the constraint curves for a successful Flash Loan attack when $n=6,800$ is shown in Fig.~\ref{fig:condition}(a). The gray area surrounded by two non-linear curves and two vertical lines shows that the favorable interval of $n_{1}$ is between 5,343.77 and 5,522.75, which can be quickly obtained by a numerical calculation. We see that $5,500$ lies in this interval. And Fig.~\ref{fig:condition}(b) shows that the maximum predictable profit one can obtain is 2,043.45 ETH if he assigns 5,343.77 out of 6,800 ETH to \textit{Compound} in step (2), which is more than $71.41+ 5,500\times(1-0.75) = 1,446.41$ ETH in the first attack transaction.

The proceed can be further maximized by tuning the amount of flash loan, $n$. The optimization process automatically executed by a numerical simulation script yields the optimized predictable profit in a few seconds. The maximum $P_{p}$ is 2,914.43 ETH with the optimal parameters set as $n=4,053$ and $n_{1}=2,596.77$, which is twice the actual $P_p$ the attacker got.

\subsubsection{Contingent Choices}
\label{subsub:contingent}
In previous analysis, we calculate the optimal predictable profit obtained under the assumption that the redemption ratio is the same as the Borrowing Exchange Rate in \textit{Compound}. By doing this we actually separate the potential gain or loss apart from the gross profit, which is resulted from the contingent choices according to step (7) and step (8).

Step (7) corresponds to a choice for the attacker to close the margin trade position and redeem the margin collateral as long as the position is not liquidated. In fact, after the attacker manipulated the price in \textit{Uniswap} in step (4), he only got about $\Delta b=51.35$ wBTC, equaling 2,002.50 ETH at a market price around 39. Such a huge loss should have already caused the short position not fully collateralized and triggered a liquidation process, which didn't occur in the real event. This was caused by a hidden bug in \textit{bZx}'s smart contract where the sanity check was skipped, which was detected by PeckShield~\footnote{https://peckshield.medium.com/bzx-hack-full-disclosure-with-detailed-profit-analysis-e6b1fa9b18fc}.

Next, by tuning $n$ and $n_1$ we see if the short position's market value will increase to yield an extra profit $P_{c1}$ denoted by Eq.~\ref{eq11-1}. The maximum value of $\Delta b \cdot 39 - \frac{l\cdot (n-n_{1})}{ocr} + (n-n_1)$ over the valid parameter space yields $-226.84$, which means unpredictable part of profit $P_{c1}$ is always 0 in this case. In the real event, the attacker didn't execute subsequent process regarding it, which is rational.

\begin{equation}
\label{eq11-1}
P_{c1} = \max{\left[0,\Delta b \cdot 39 - \frac{l\cdot (n-n_{1})}{ocr} + (n-n_1)\right]}
\end{equation} 
%
%
%


As for step (8), a reasonable strategy is to keep cETH until the price of wBTC decreases, expecting a positive extra profit $P_{c2}$. If the expected price of wBTC is going to increase, it will be better to redeem the collateral instantly since it will depreciate. In the real event, the price of wBTC increased from $p=36.48$ ETH to $p_m=39.08$ ETH. The attacker might have noticed it and started to redeem the collateral in the subsequent transactions. To minimize the impact it may bring to the market, the attacker finished the redemption process by sending multiple small-amount transactions. The process extended over several \textit{Ethereum} blocks and the redemption ratio was not predictable. Part of collateral even had been liquidated by third parties. In a word, $P_{c2}$ is unpredictable and can be negative when the collaterized assets go through a depreciation. That's why the ultimate gross profit in this event, about $1244.108$ ETH as shown in Fig.~\ref{fig:bzxpump}, is less than the predictable profit $1446.41$ ETH.  

\subsection{Summary}
From this attack, the composability of DeFi is demonstrated by Flash Loan transaction vividly. Flash Loan attack strategy connects different \textit{money legos} such as Flash Loan, decentralized lending platform, margin trade platform, exchange router and AMM, and the margin trade platform itself is composed of lower-level modules. Flash Loan also offered a huge sum of money to help the attacker conduct such a pump-and-dump attack strategy at a negligible error cost.

The strategy will not succeed if two prerequisites are missing. 

\begin{itemize}

\item First, there is an AMM where price can be predicted and manipulated easily. In this case,  \textit{Uniswap} is the target, where price is predicted by a constant product formula and converges to the external market price relying on on-chain arbitrage bot. By making a single large-volume transaction one can pump the price of the reduced asset in a swap pair to an extremely high level. 

\item Second, the atomic property of Flash Loan transaction works like a shelter against external arbitrageur who may detect and reduce the price spread instantly. The arbitrage spread is manually created and then made use of by the attacker in one go.

\end{itemize}

Similar to the analysis in ~\cite{attackQin}, the victim in this attack is protocol \textit{bZx}, which encountered a loss of about $5,637.62 - 1,300 - 51.346\times 39.08 = 2,331.02$ ETH since the attacker's margin trade position was depreciated dramatically. The liquid providers in \textit{Uniswap} encountered a great impermanent loss due to the large price deviation, which could be taken advantage of by arbitrageurs to get a profit about $1,161.26$ ETH~\footnote{After step (4), there were 138.76 wBTC and 1565.21 ETH in \textit{Uniswap}'s wBTC asset pool and the price of wBTC is dumped to be 11.28 ETH. To pull the price back to the market price, say, 39.08, the arbitrageur swaps 1,348.19 ETH for 64.21 wBTC with 29.72wBTC as an arbitrage gain.}. The attacker finally grabbed $1244.11$ ETH (\$330K) into his own address. It is noted that the loss value of \textit{bZx} roughly equals the sum of the arbitrage gain in \textit{Uniswap} and the attacker's profit. And the proceeds could be doubled if the parameters were set optimally. 

\section{The Trends of Flash Loan Attacks}
\label{sec:trends}

We conduct a thorough study on other Flash Loan attacks as listed in Table~\ref{table:events}. \textit{Flashots} can be found in Appendix~\ref{appendix}. From these \textit{Flashots}, we can summarize a few development trends of the Flash Loan attacks.

\begin{itemize}
\item The proceeds grabbed by the exploiters increased by a scale. This is partially due to the expanding of the capital scale in DeFi market itself. 
\item While some of the attacks include a series of sophisticate steps in one strike, some others show a trend to repeat simple steps to accumulate profits.
\item While most of the attack targets are AMM, the role AMM plays is different from each other. In most Flash Loan attack events, AMM acts as an oracle providing prices to other DeFi protocols, so that attackers shift their target to the oracles and then they can manipulate the exchange or borrowing ratios of certain tokens that relying on the oracle.
\item While some events happened due to the technical bugs of smart contracts such as reentrancy attacks and some of the oracle manipulation attacks like Warp Finance Attack~\footnote{https://www.coindesk.com/warp-finance-suffers-possible-8m-flash-loan-attack}, the rest events show vulnerabilities emerging from the DeFi systems themselves.
\end{itemize}

Without doubt, Flash Loan attacks will continue to occur in the future. To control risks, some protocols introduced maximum slippage checks in an AMM swap, yet attackers were still capable to perform manipulations within limitation and many a little makes a mickle, such as in Harvest Attack~\footnote{https://www.coindesk.com/harvest-finance-24m-attack-triggers-570m-bank-run-in-latest-defi-exploit} as shown in Fig.~\ref{fig:harvestusdc} and Fig.~\ref{fig:harvestusdt}. Another solution is to make sure that the buying and selling operations of an asset must be executed in different transactions or even blocks so that the attacker cannot repay flash loan in one transaction, but this is at the expense of normal users' experience. 

From another perspective, Flash Loan attacks have contributed a lot to the discovery of DeFi protocols' vulnerabilities. We could see that similar strategies were used by the attackers, such as Cheese Bank Attack and Warp Finance Attack~\footnote{Both of the attackers grabbed their proceeds by manipulating the amount of an asset in AMM that is a critical parameter used by the price oracle.} as shown in Fig.~\ref{fig:cheese} and Fig.~\ref{fig:warp}, which can be completely avoided based on the lessons learned. These events warned us that protocols should be designed more seriously in against both of the attackers and deal-hunters. \texttt{Flashot} is designed to be a helpful tool for this task.
\section{Conclusion and Discussion}
\label{sec:conclusion}
In this paper, we studied an unprecedented lending tool, Flash Loan, which is a double-edged sword to be used either to raise capital efficiency during normal financial operations such as liquidation, or to conduct a subtly designed Flash Loan attack.

Since transactions can only be executed one by one and the update time of blockchain state is discrete, the Flash Loan transaction makes an exclusive and definite change of the blockchain's world state as if DeFi systems are stopped running during its execution. It somewhat realizes a scene depicted by \textit{Time Dilation} theory proposed by Albert Einstein. Flash Loan transaction is like a rocket flying across the blockchain world at the speed of light, and in the view from the blockchain world, time is dilated. Thus it's interesting to call Flash Loan attack as \textit{Time Dilation} attack, where Flash Loan transaction sender can apply magic changes to the blockchain's world state with no one else can interfere. 

During the magic changes, assets borrowed from the flash loan pools shuttle smartly among different DeFi protocols. \texttt{Flashot} we proposed in this paper can be used to illustrate the asset flows intertwined with smart contracts in a standard way, which is like taking a snapshot to capture the running process of the Flash Loan transaction. In the future, the modules assembled by DeFi systems may become hundreds and thousands. At that time such a standardized tool will show more power.

The features caught by \texttt{Flashot} can help form a comprehensive understanding about the key components to be improved in order to make a more robust and efficient DeFi system. In bZx Pump Attack, we see that currently the core target of Flash Loan is liquidity pools of AMM, especially those swap pairs with low liquidity. To some extend, the size of the liquidity pool will become the "moat" for AMM to resist Flash Loan attacks. Driven by such a factor, the Flash Loan attack may facilitate the optimization and consolidation of the AMM liquidity pools. AMMs whose liquidity pools have not reached the "critical scale" may be absorbed by larger AMMs.

We also realize a fusion effect of the DeFi ecosystem on \textit{Ethereum}. Before frequent Flash Loan attack incidents happened, DeFi's \textit{money legos} had just shown a limited composability. As the DeFi ecosystem has already developed a scale on \textit{Ethereum}, it offers a diverse platform for Flash Loan borrowers to connect imaginative composite of \textit{money legos}. On the other hand, Flash Loan attack shows a great potential to speed up the construction and troubleshooting of DeFi systems in a positive way. These all form a virtuous cycle helping transform the rudimentary DeFi systems to decentralized financial infrastructures that can realize more sophisticated functions. And the difficulty of migrating DeFi's building blocks from \textit{Ethereum} to other public blockchains is increasing because the latter does not have such an ecosystem with comparable scale.

Moreover, Flash Loan transaction provides a new solution for reducing frictions in the financial systems. One do not need to hold the principal to participate and the liquidity risk is greatly reduced. Since the revertion feature cannot be established without blockchain technology, it also demonstrates that in addition to solving credit problems by providing atomic transactions, blockchain technology can also greatly improve the capital efficiency through Flash Loan. In future, competition in the financial industry may just have to focus on how to improve the capabilities on advanced modeling and system design in terms of price discovery. 

To conclude, We look forward to a next-generation of financial industry powered by highly efficient automatic risk and profit detection systems based on the blockchain.

\section*{Acknowledgments}

We thank Dr. Xiao Feng for insightful discussions about the development of Flash Loan and the significant impact it may bring to DeFi ecosystem.

%

\bibliographystyle{plain}
\bibliography{\jobname}

\begin{thebibliography}{10}

\bibitem{uniswap}
Hayden Adams, Noah Zinsmeister, and Dan Robinson.
\newblock Uniswap v2 core.
\newblock 2020.
\newblock \url{https://uniswap.org/whitepaper.pdf}.

\bibitem{vitalik2013}
Vitalik Buterin.
\newblock Ethereum: A next-generation smart contract and decentralized
  application platform.
\newblock 2013.
\newblock \url{https://ethereum.org/en/whitepaper/}.

\bibitem{JBVIchen}
Yan Chen and Cristiano Bellavitis.
\newblock Blockchain disruption and decentralized finance: The rise of
  decentralized business models.
\newblock {\em Journal of Business Venturing Insights}, 13:e00151, 2020.
\newblock \url{https://doi.org/10.1016/j.jbvi.2019.e00151}.

\bibitem{thesis}
Florian Gronde.
\newblock {\em Flash Loans and Decentralized Lending Protocols: An In-Depth
  Analysis}.
\newblock PhD thesis, University of Basel, 7 2020.
\newblock
  \url{https://wwz.unibas.ch/fileadmin/user_upload/wwz/00_Professuren/Schaer_DLTFintech/Lehre/MA_Florian_Gronde_Flashloans-ohne_Appendix.pdf}.

\bibitem{crisis}
Lewis Gudgeon, Daniel Perez, Dominik Harz, Benjamin Livshits, and Arthur
  Gervais.
\newblock The decentralized financial crisis.
\newblock {\em 2020 Crypto Valley Conference on Blockchain Technology (CVCBT)},
  2020.
\newblock \url{https://doi.org/10.1109/CVCBT50464.2020.00005}.

\bibitem{coingeckobook}
Darren Lau, Daryl Lau, Sze~Jin Teh, Kristian Kho, Erina Azmi, Lee TM, and Bobby
  Ong.
\newblock {\em How to DeFi}.
\newblock Independently published, 1st edition, 2020.
\newblock \url{https://www.amazon.com/How-DeFi-CoinGecko/dp/B0884B51KG}.

\bibitem{xuefeng}
Xuefeng Li, Xiaochuan Wu, Xin Pei, and Zhuojun Yao.
\newblock Tokenization: Open asset protocol on blockchain.
\newblock {\em 2019 IEEE 2nd International Conference on Information and
  Computer Technologies (ICICT)}, 2019.
\newblock \url{https://doi.org/10.1109/INFOCT.2019.8711021}.

\bibitem{bitcoin}
Satoshi Nakamoto.
\newblock Bitcoin: A peer-to-peer electronic cash system.
\newblock 2008.
\newblock \url{https://bitcoin.org/bitcoin.pdf}.

\bibitem{attackQin}
Kaihua Qin, Liyi Zhou, Benjamin Livshits, and Arthur Gervais.
\newblock Attacking the defi ecosystem with flash loans for fun and profit.
\newblock {\em arXiv preprint arXiv:2003.03810v2}, 2020.
\newblock \url{https://arxiv.org/abs/2003.03810}.

\bibitem{tubiblio111410}
Michael Rodler, Wenting Li, Ghassan~O. Karame, and Lucas Davi.
\newblock Sereum: Protecting existing smart contracts against re-entrancy
  attacks.
\newblock In {\em Proceedings of 26th Annual Network \& Distributed System
  Security Symposium (NDSS)}, 2019.
\newblock \url{http://tubiblio.ulb.tu-darmstadt.de/111410/}.

\bibitem{tokenization2}
Jakob Roth, Fabian Sch\"{a}r, and Aljoscha Sch\"{o}pfer.
\newblock The tokenization of assets: Using blockchains for equity
  crowdfunding.
\newblock {\em Available at SSRN}, 2019.
\newblock \url{https://ssrn.com/abstract=3443382}.

\bibitem{schar}
Fabian Sch\"{a}r.
\newblock Decentralized finance: On blockchain- and smart contract-based
  financial markets.
\newblock {\em Available at SSRN}, 2020.
\newblock \url{https://ssrn.com/abstract=3571335}.

\bibitem{tokenization1}
Yifeng Tian, Yuanxin Zhang, R.~Edward Minchin, Ashish Asutosh, and Congwen Kan.
\newblock An innovative infrastructure financing instrument: Blockchain-based
  tokenization.
\newblock {\em Construction Research Congress 2020: Infrastructure Systems and
  Sustainability}, 2020.
\newblock \url{https://doi.org/10.1061/9780784482858.079}.

\bibitem{flashWang}
Dabao Wang, Siwei Wu, Ziling Lin, Lei Wu, Xingliang Yuan, Yajin Zhou, Haoyu
  Wang, and Kui Ren.
\newblock Towards understanding flash loan and its applications in defi
  ecosystem.
\newblock {\em arXiv preprint arXiv:2010.12252v1}, 2020.
\newblock \url{https://arxiv.org/abs/2010.12252}.

\bibitem{werner2021sok}
Sam~M. Werner, Daniel Perez, Lewis Gudgeon, Ariah Klages-Mundt, Dominik Harz,
  and William~J. Knottenbelt.
\newblock Sok: Decentralized finance (defi).
\newblock {\em arXiv preprint arXiv:2101.08778v1}, 2021.
\newblock \url{https://arxiv.org/abs/2101.08778}.

\bibitem{GavinWood}
Gavin Wood.
\newblock Ethereum: A secure decentralised generalised transaction ledger.
\newblock 2020.
\newblock \url{https://ethereum.github.io/yellowpaper/paper.pdf}.

\bibitem{aave}
wow@aave.com.
\newblock Aave protocol whitepaper.
\newblock 2020.
\newblock
  \url{https://github.com/aave/aave-protocol/blob/master/docs/Aave\_Protocol\_Whitepaper\_v1\_0.pdf}.

\end{thebibliography}

\appendix
\section{Appendices}
\label{appendix}
\renewcommand\thefigure{\Alph{section}\arabic{figure}}
\setcounter{figure}{0}
\begin{figure*}[h]
\centering
\includegraphics[width=1\textwidth]{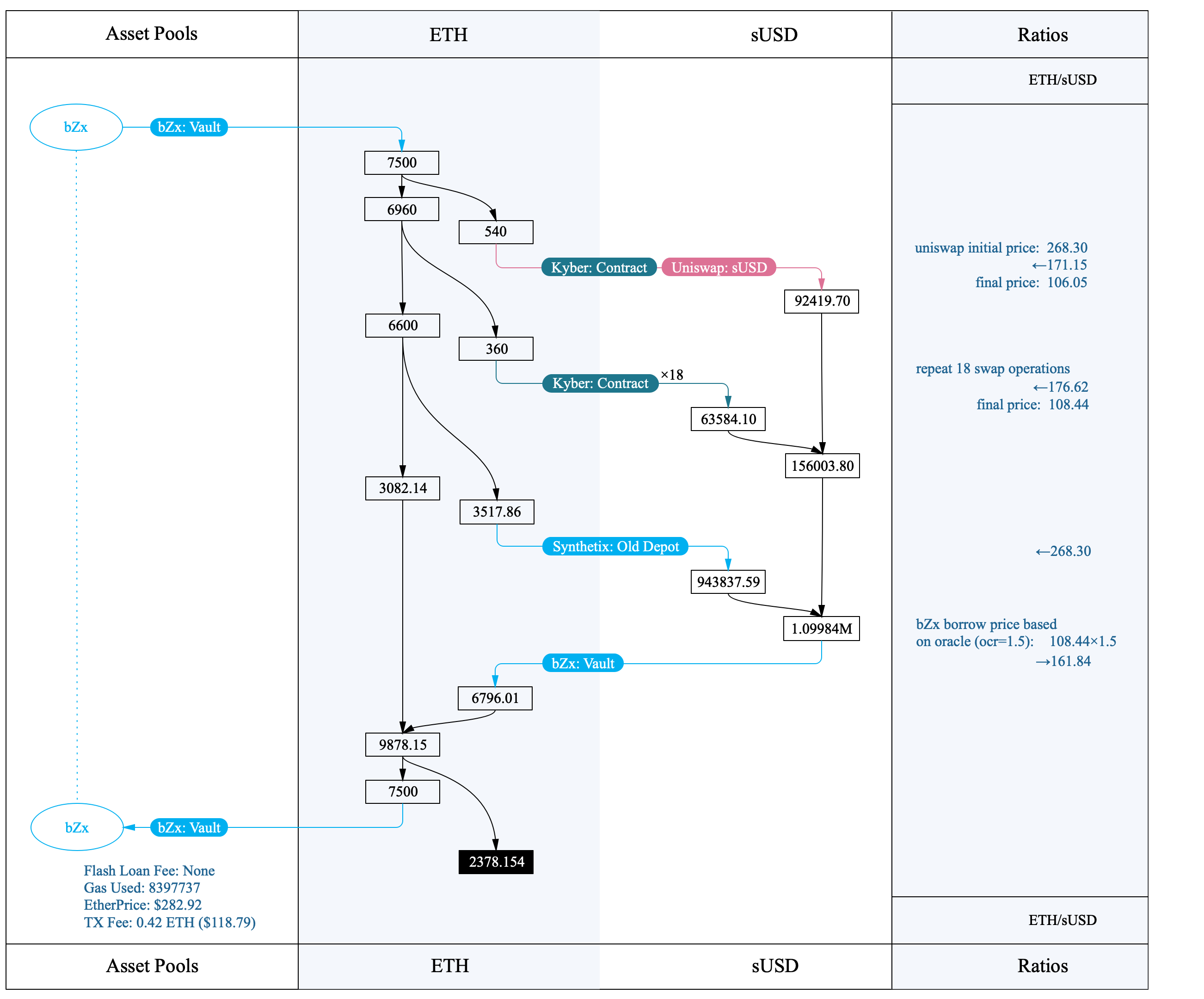}
\caption{\textit{Flashot} of bZx Oracle Attack. The borrowing rate is based on an oracle depending on the prices provided by \textit{Uniswap} and \textit{Kyber}, which was manipulated by the attacker in this event. The whole diagram presents the process of a single Flash Loan transaction, where 18 swap operations interacted with \textit{Kyber}'s smart contract are merged for simplicity. Txhash: 0x762881b07feb63c436dee38edd4ff1f7a74c33091e534af56c9f7d49b5ecac15.}
\label{fig:bzxoracle}
\end{figure*}

\begin{figure*}[h]
\centering
\includegraphics[width=1\textwidth]{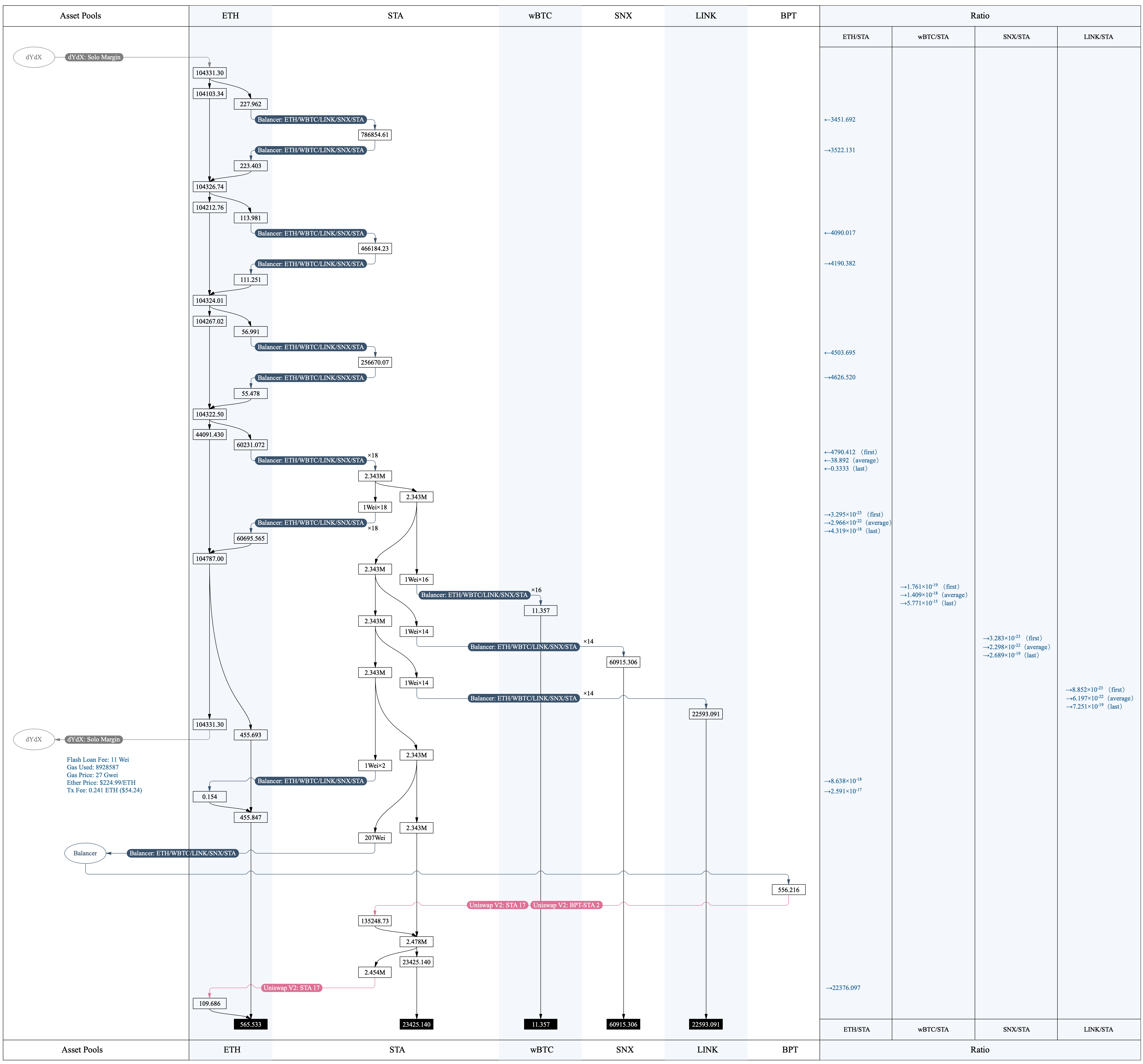}
\caption{\textit{Flashot} of Balancer Attack. The attacker drained up Balancer's asset pool by manipulating the price of a deflationary token called STA. Since in each swap operation 1\% of STA will be burned as a swap fee, the attacker swapped it frequently with other 4 tokens and then the amount of STA was dramatically reduced and STA's price soared sharply. Txhash: 0x013be97768b702fe8eccef1a40544d5ecb3c1961ad5f87fee4d16fdc08c78106.}
\label{fig:balancer}
\end{figure*}

\begin{figure*}[h]
\centering
\includegraphics[width=1\textwidth]{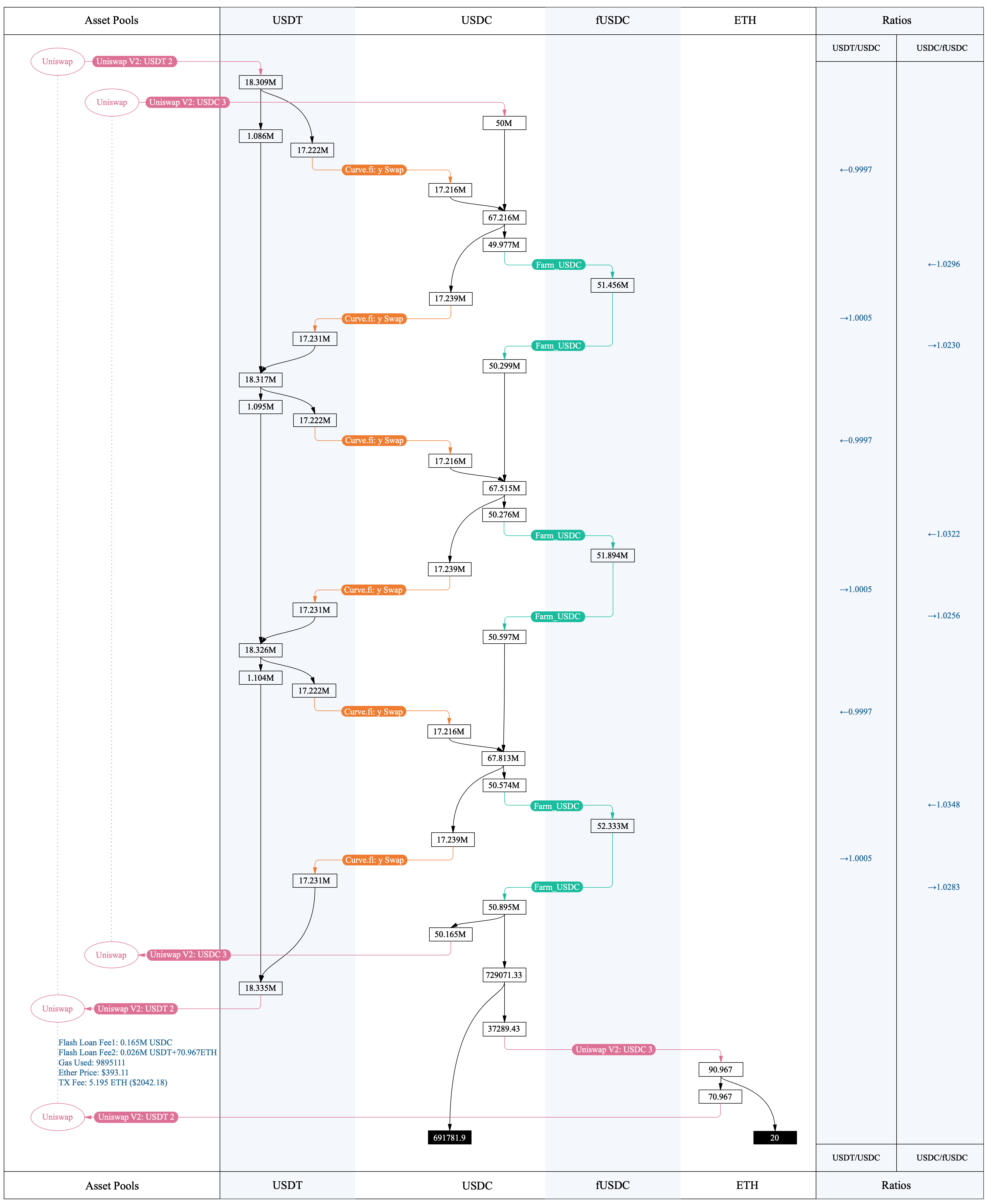}
\caption{\textit{Flashot} of the first Flash Loan transaction targeting at \textit{Harvest}'s fUSDC pool in Harvest Attack, which was repeated by 16 times to grab a total of 14 million USDC and 340 ETH. Txhash: 0x35f8d2f572fceaac9288e5d462117850ef2694786992a8c3f6d02612277b0877.}
\label{fig:harvestusdc}
\end{figure*}

\begin{figure*}[h]
\centering
\includegraphics[width=1\textwidth]{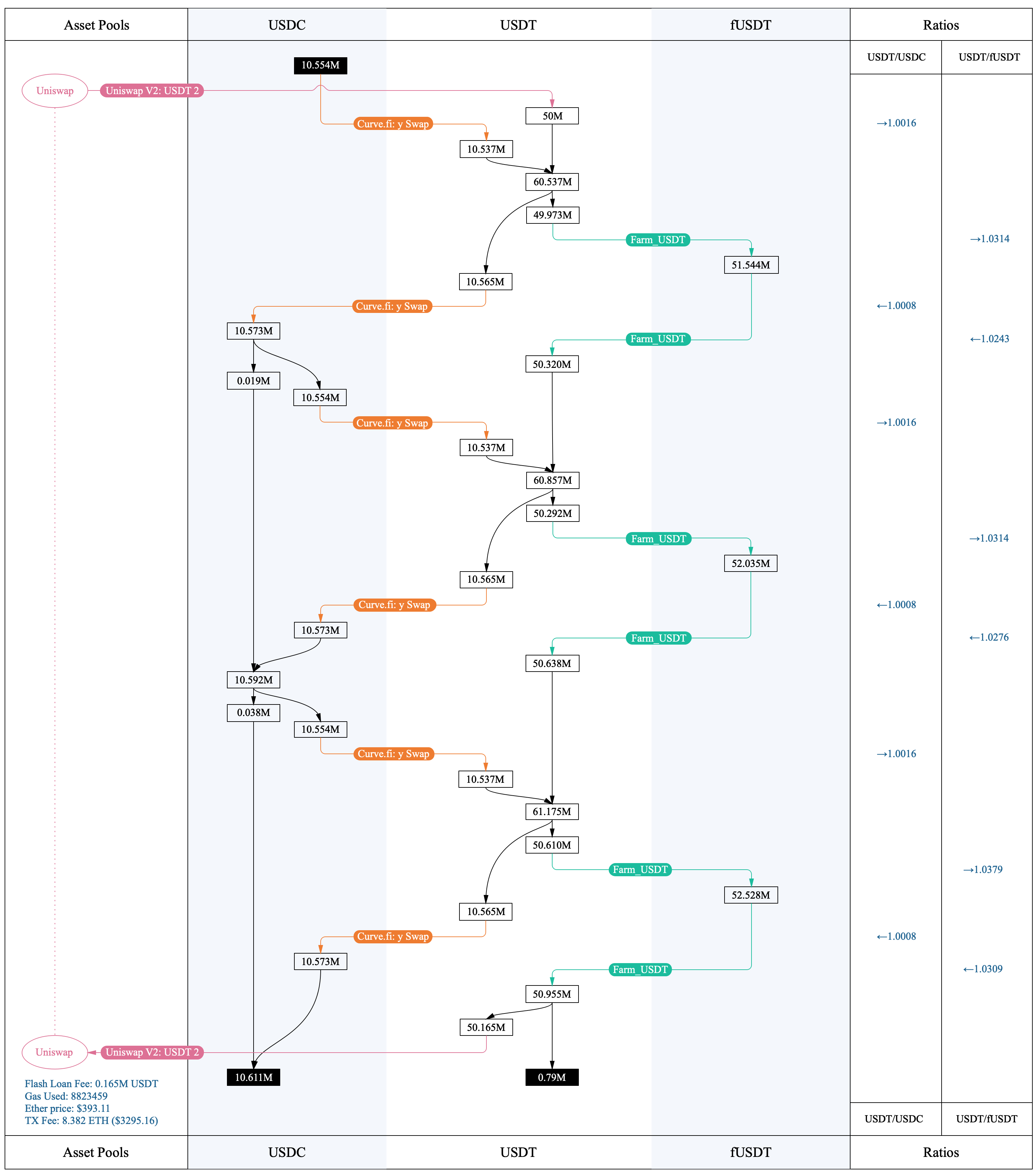}
\caption{After attacking fUSDC pool as shown in Fig.~\ref{fig:harvestusdc}, the attacker repeated a similar strategy by 13 times targeting at \textit{Harvest}'s fUSDT pool and grabbed a total of 11.7 million USDT and 0.76 million USDC. Here we show the \textit{flashot} of the first Flash Loan transaction targeting at \textit{Harvest}'s fUSDT pool in this event. Some proceeds out of previous attack transactions were used in the subsequent attacks. Txhash: 0x0fc6d2ca064fc841bc9b1c1fad1fbb97bcea5c9a1b2b66ef837f1227e06519a6.}
\label{fig:harvestusdt}
\end{figure*}

\begin{figure*}[p]
	\begin{adjustbox}{addcode={
		\begin{minipage}{\width}}{
			\caption{\textit{Flashot} of Cheese Bank Attack. The borrowing rate is provided by an oracle, which was manipulated by the attacker through increasing the amount of \textit{Uniswap}'s CHEESE 2 pool's ETH. Txhash: 0x600a869aa3a259158310a233b815ff67ca41eab8961a49918c2031297a02f1cc.}
			\label{fig:cheese}
		\end{minipage}},rotate=90,center}
		\includegraphics[width=1\textheight]{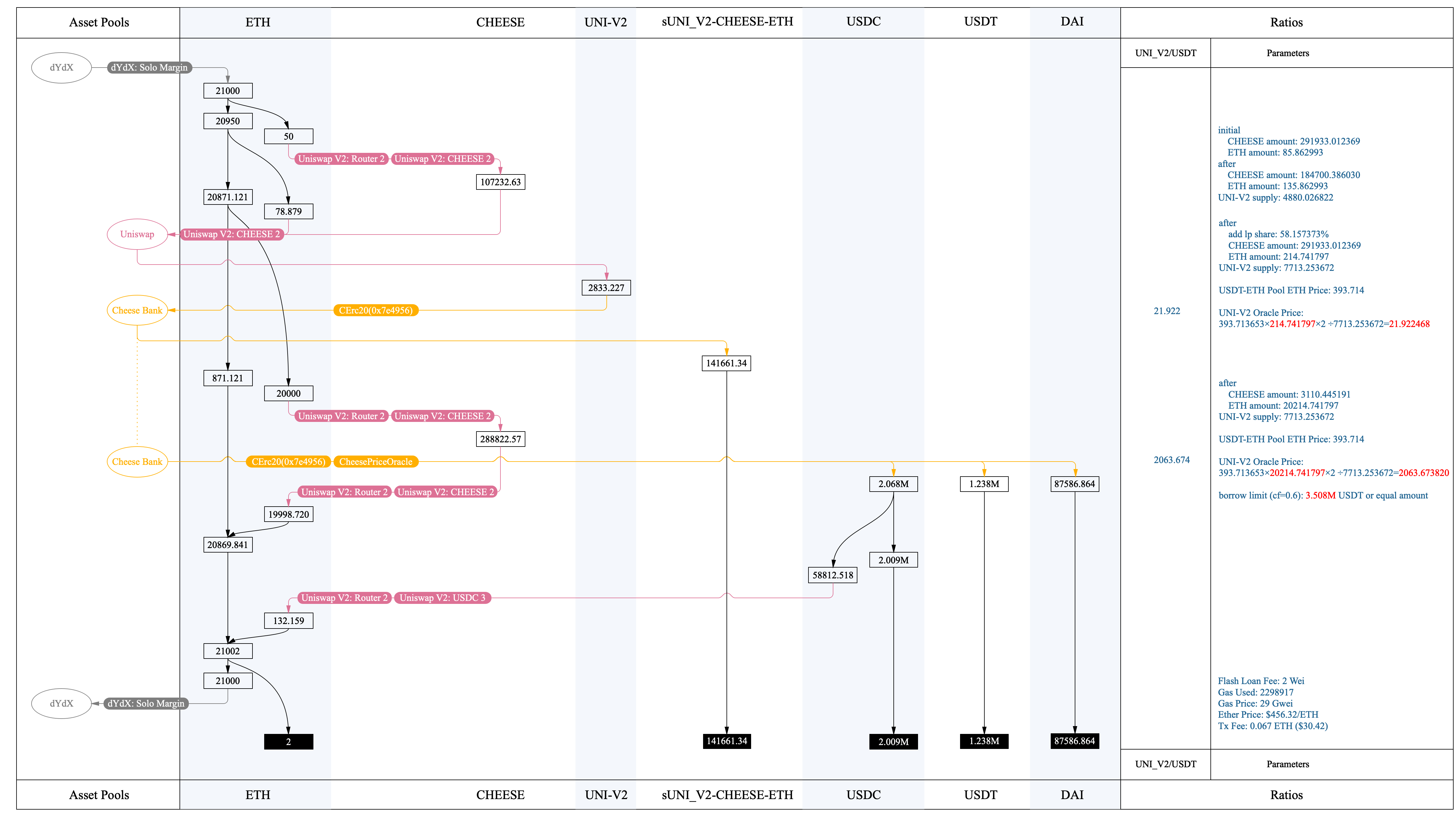}
	\end{adjustbox}
\end{figure*}

\begin{figure*}[h]
\centering
\includegraphics[height=0.9\textheight]{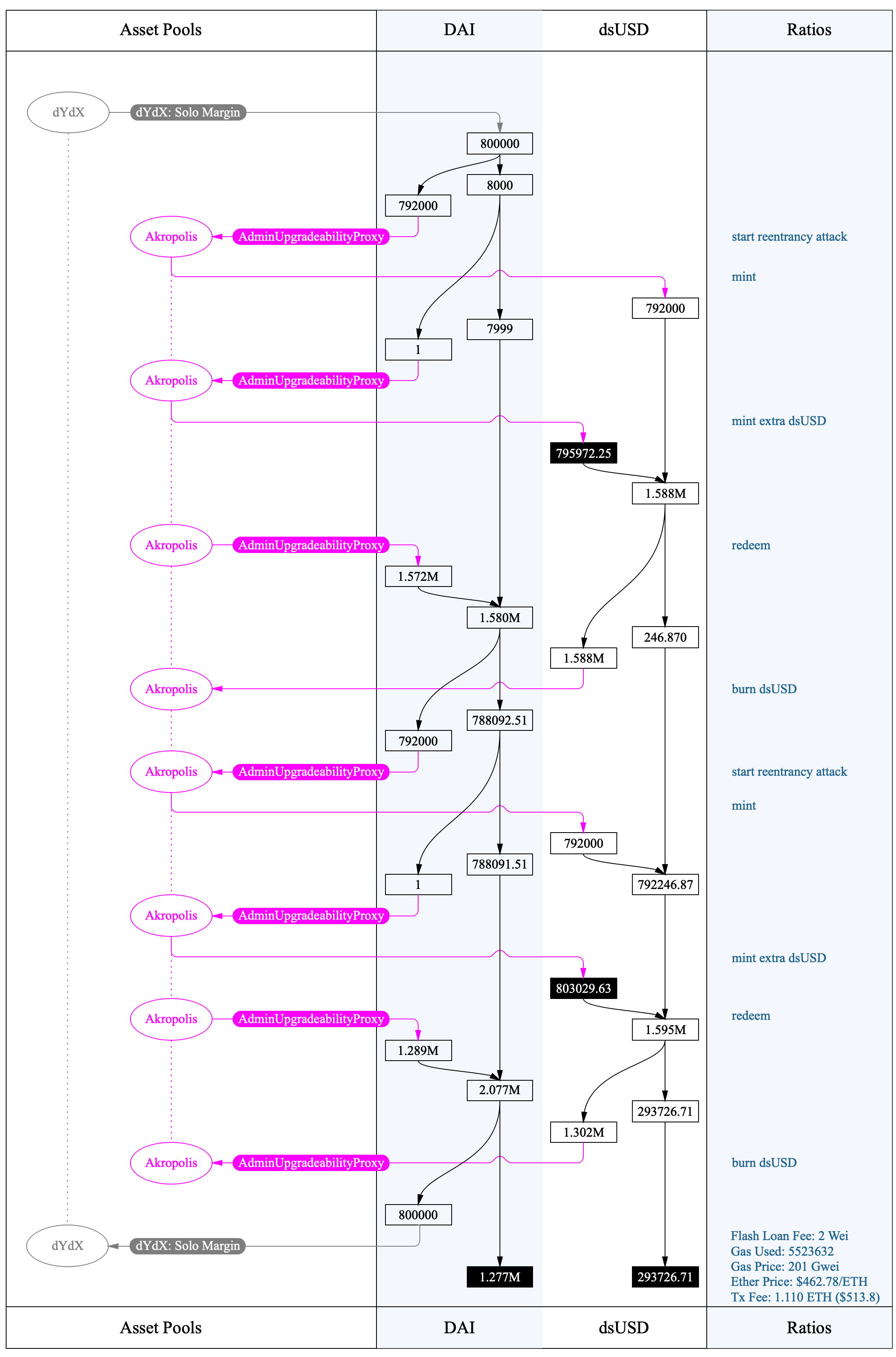}
\caption{\textit{Flashot} of the first Flash Loan transaction exploited in a reentrancy attack targeted at \textit{Akropolis}. A large number of dsUSD was minted without being backed by any collateral. The attacker repeated the same strategy in another 16 transactions and got a total of 2.04 million DAI and 0.31 million dsUSD. Txhash: 0xddf8c15880a20efa0f3964207d345ff71fbb9400032b5d33b9346876bd131dc.}
\label{fig:akropolis}
\end{figure*}

\begin{figure*}[p]
	\begin{adjustbox}{addcode={
		\begin{minipage}{\width}}{
			\caption{\textit{Flashot} of Value.DeFi Attack. The price of textit{Value.DeFi}'s pool token mvUSD is fed by \textit{Curve} as an oracle. The attacker minted  mvUSD at a normal price and then manipulated the price of 3Crv by swapping a large amount of stables coins in \textit{Curve} 's DAI/USDC/USDT pool, which pumped the price of mvUSD to redeem more 3Crv. Txhash: 0x46a03488247425f845e444b9c10b52ba3c14927c687d38287c0faddc7471150a.}
			\label{fig:valuedefi}
		\end{minipage}},rotate=90,center}
		\includegraphics[width=1\textheight]{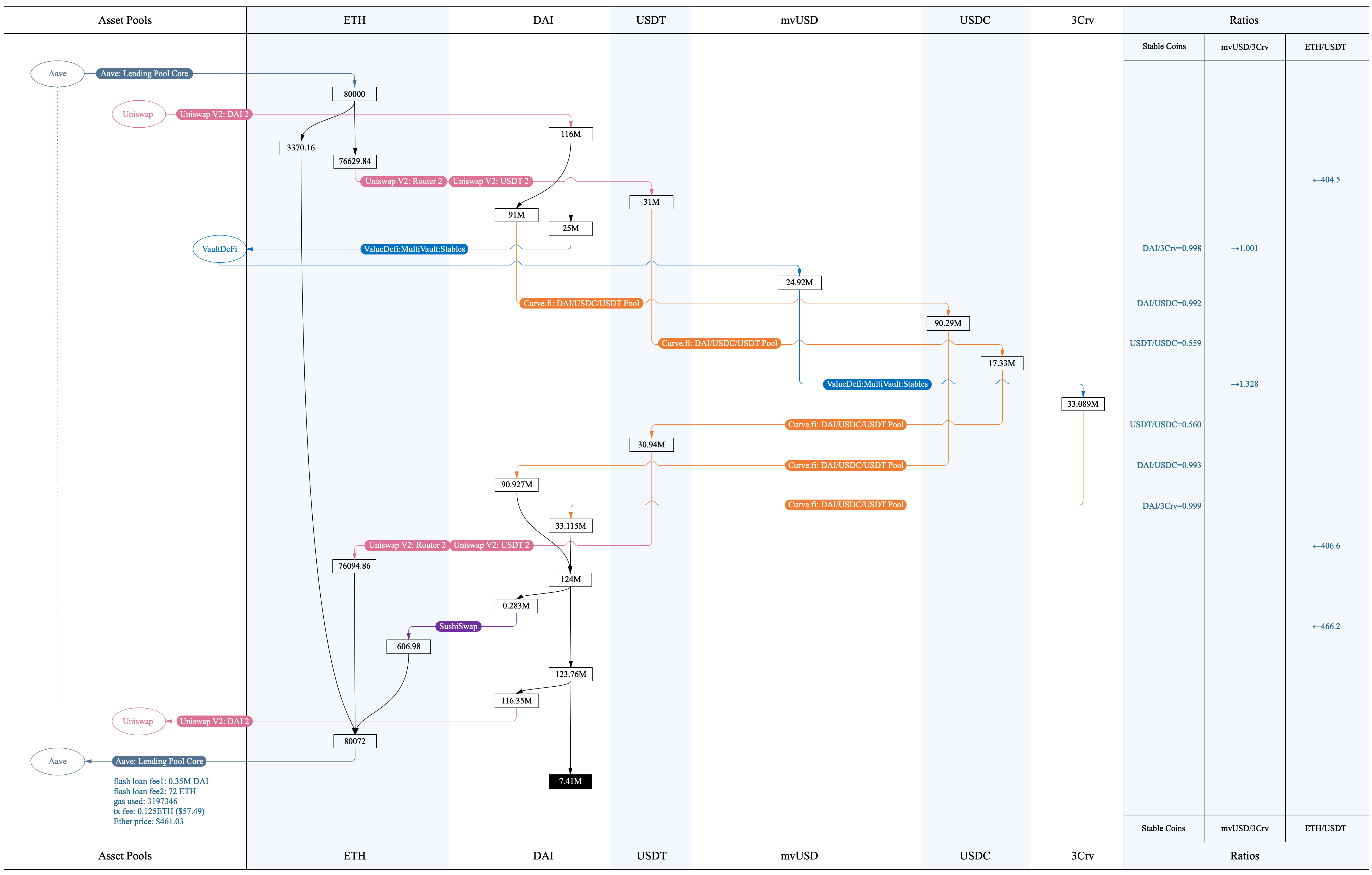}
	\end{adjustbox}
\end{figure*}

\begin{figure*}[h]
\centering
\includegraphics[height=0.9\textheight]{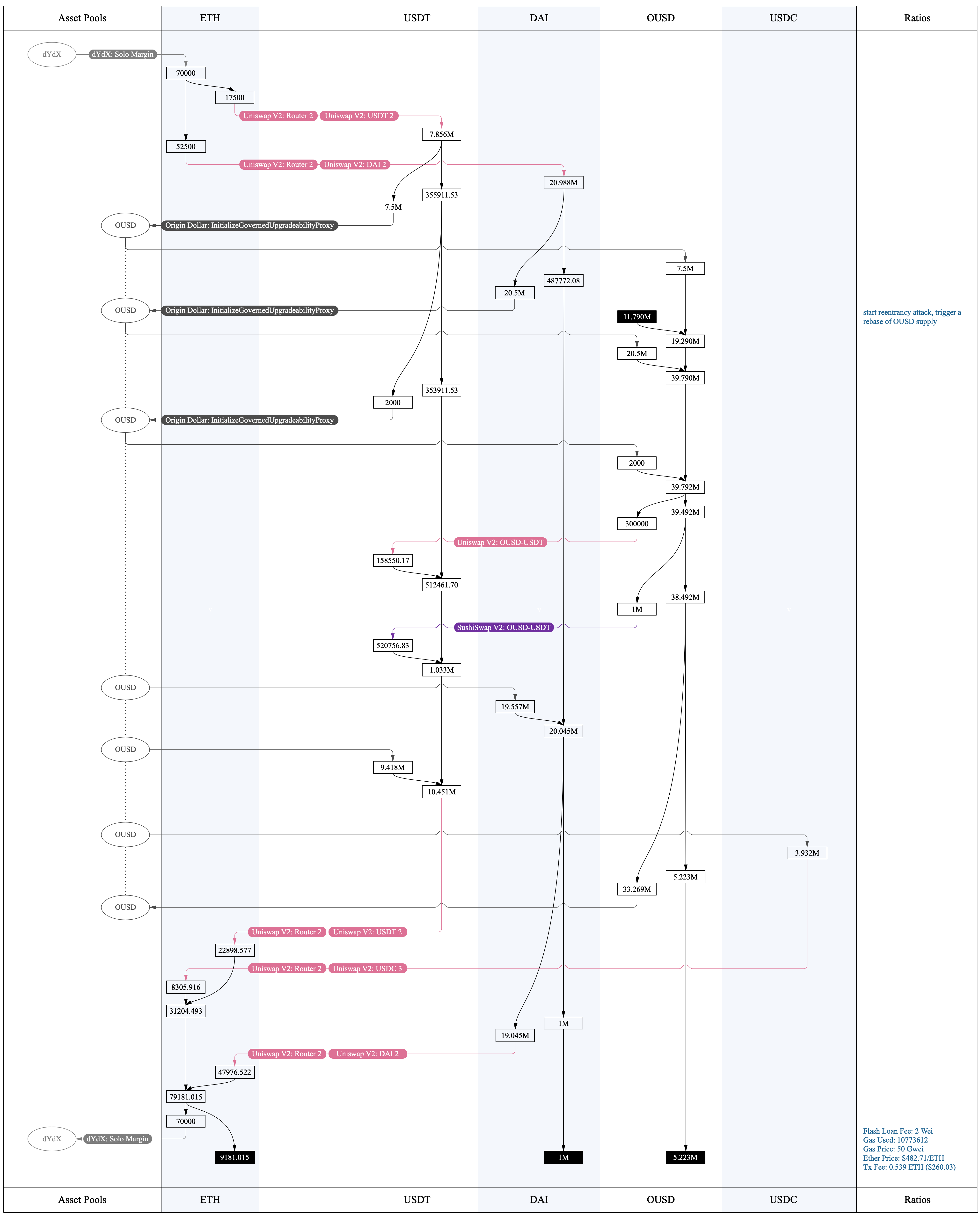}
\caption{\textit{Flashot} of the Flash Loan transaction exploited in a reentrancy attack targeted at \textit{OUSD}. The balances of all user's OUSD were rebased and the attacker got a total of 9181 ETH, 1 million DAI and 5.223 million OUSD. The attacker sent 11 subsequent transactions to redeem OUSD. Txhash: 0xe1c76241dda7c5fcf1988454c621142495640e708e3f8377982f55f8cf2a8401.}
\label{fig:ousd1}
\end{figure*}

\begin{figure*}[h]
\centering
\includegraphics[width=1\textwidth]{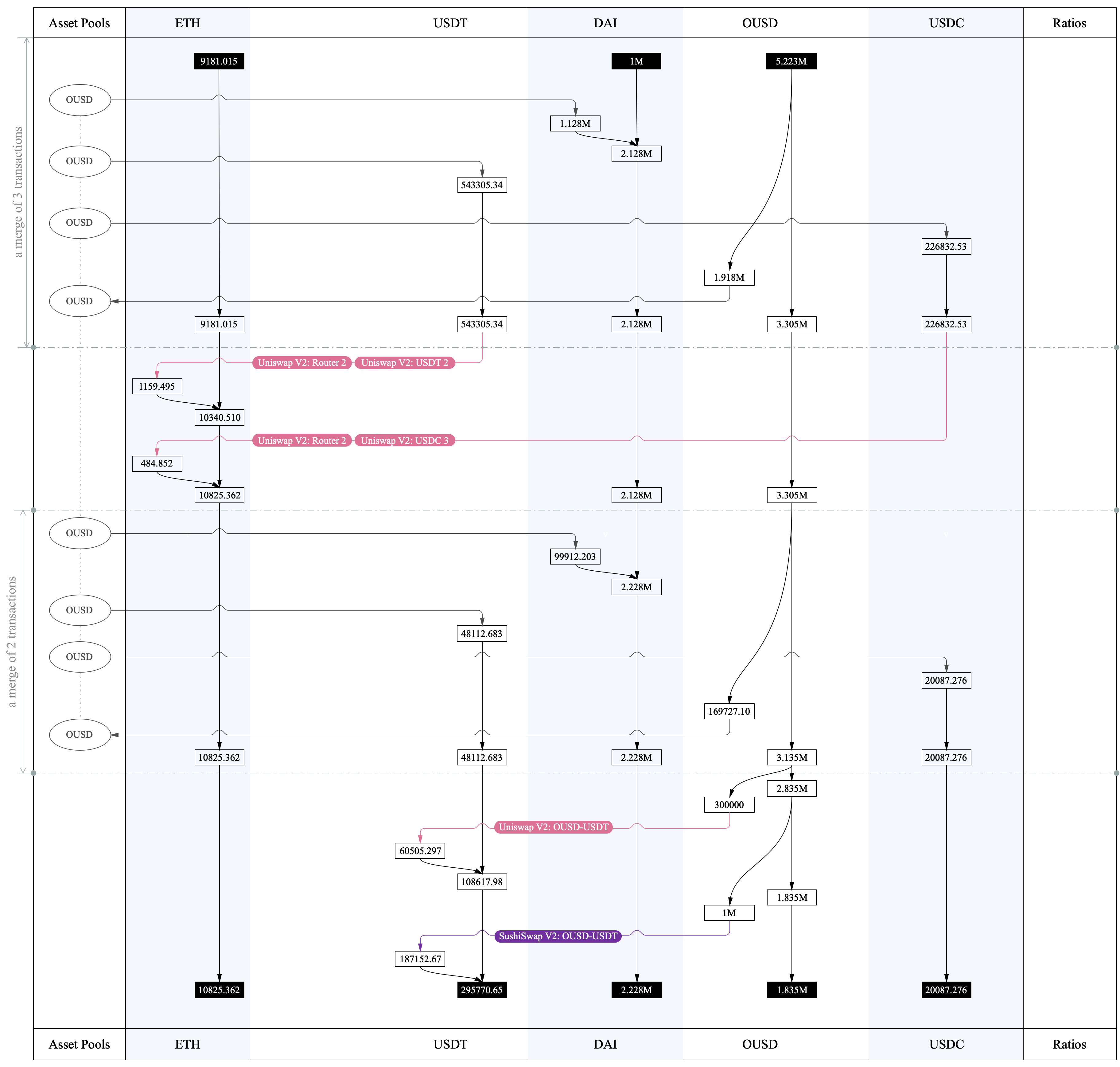}
\caption{\textit{Flashot} of subsequent transactions in the reentrancy attack targeted at \textit{OUSD}. Some transactions with a same pattern were merged for simplicity.}
\label{fig:ousd2}
\end{figure*}

\begin{figure*}[h]
\centering
\includegraphics[width=1\textwidth]{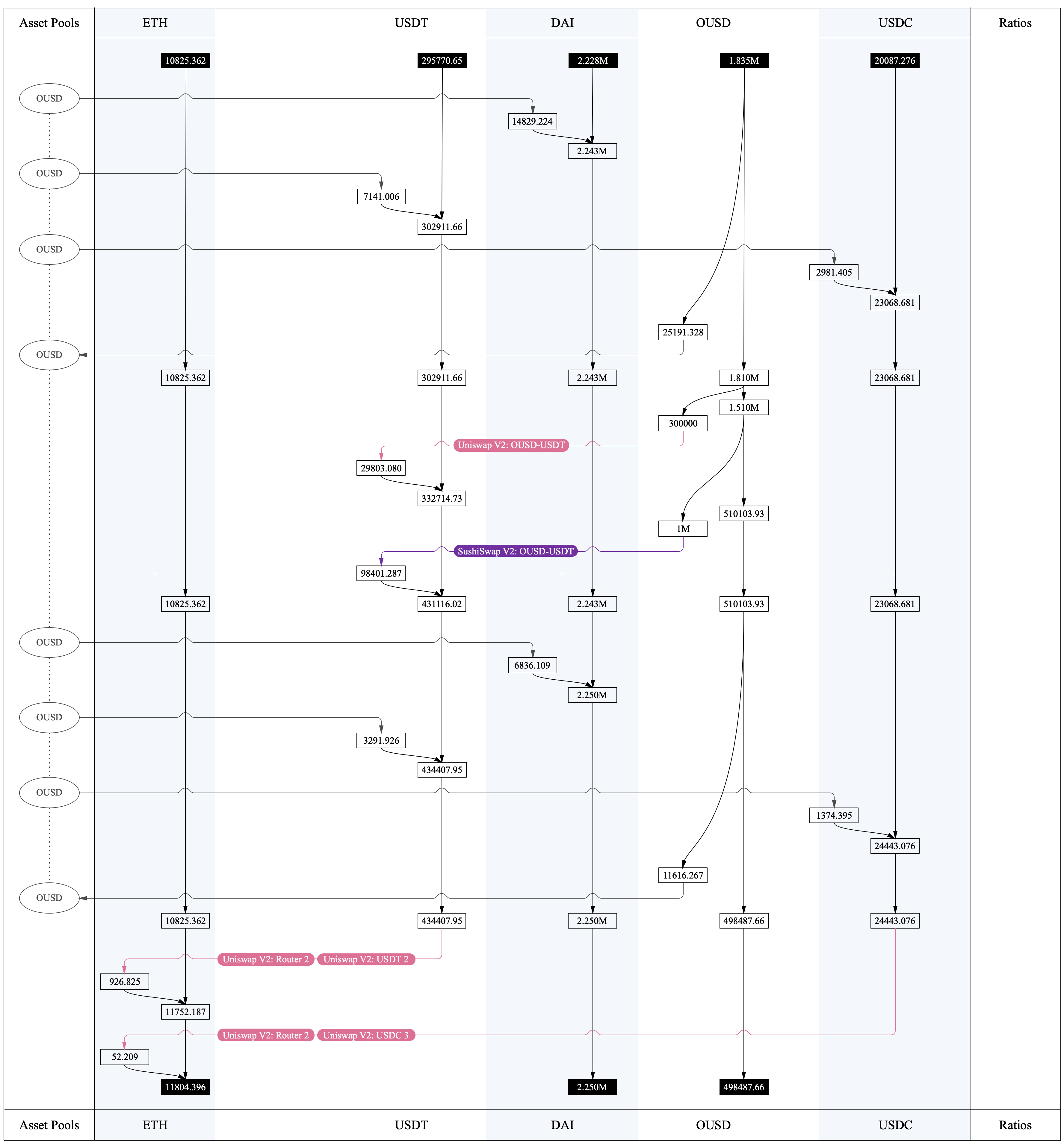}
\caption{\textit{Flashot} of subsequent transactions in the reentrancy attack targeted at \textit{OUSD} as a continuation of Fig.~\ref{fig:ousd2}.}
\label{fig:ousd3}
\end{figure*}

\begin{figure*}[h]
\centering
\includegraphics[width=1\textwidth]{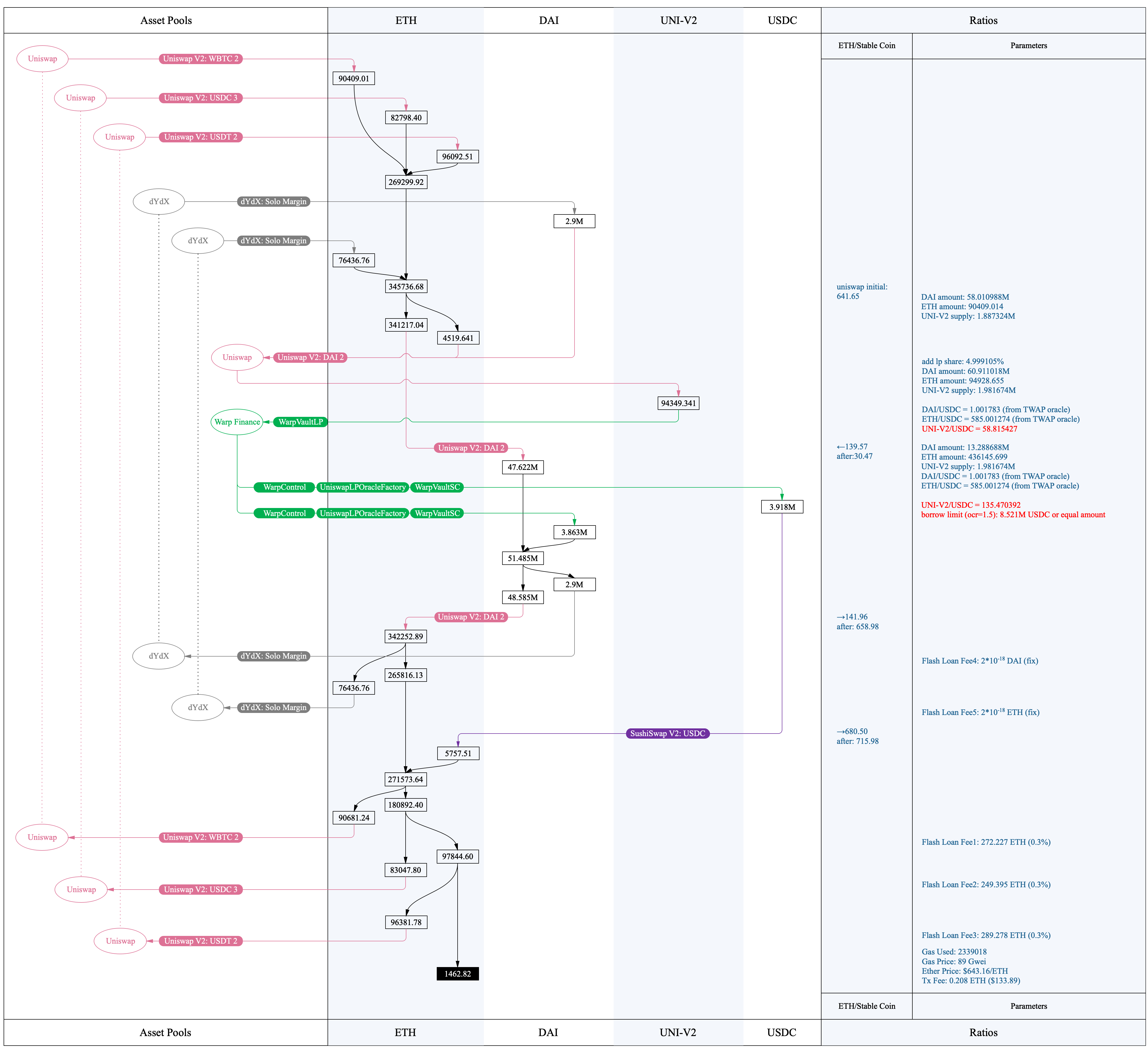}
\caption{\textit{Flashot} of Warp Finance Attack in a single Flash Loan transaction. The attacker exploited a calculation bug in Warp Finance's oracle, where the price of \textit{Uniswap}'s pool token UNI-V2 was calculated by $\frac{n_{ETH}\cdot p_{ETH} + n_{DAI}\cdot p_{DAI}}{total supply}$. While prices $p_{ETH}$ and $p_{DAI}$ were normally provided by \textit{Uniswap}'s official time-waited-average-price (TWAP) oracle, the amount $n_{ETH}$ and $n_{DAI}$ were manipulated to make the collateral UNI-V2 appreciate dramatically so that the attacker could borrow more. Txhash: 0x8bb8dc5c7c830bac85fa48acad2505e9300a91c3ff239c9517d0cae33b595090.}
\label{fig:warp}
\end{figure*}

\end{document}